\documentclass{aa}  

\usepackage{graphicx}
\usepackage{txfonts}
\usepackage{multicol}        % Multi-column entries in tables
\usepackage{multirow}
\usepackage{soul}
\usepackage{natbib}
\bibpunct{(}{)}{;}{a}{}{,} % to follow the A&A style

\usepackage[colorlinks=true, urlcolor=blue, colorlinks=true, citecolor=blue, linkcolor=blue]{hyperref}

\begin{document} 

   \title{The redshift evolution of galactic bar pattern speed in TNG50}
   
   \author{Asiyeh Habibi \inst{1}\thanks{\email{asiyeh.habibi@alumni.um.ac.ir}} \and Mahmood Roshan
          \inst{1,}\inst{2} \and Mohammad Hosseinirad \inst{2} \and Habib Khosroshahi \inst{2} \and J. A. L. Aguerri \inst{3,}\inst{4} \and Virginia Cuomo \inst{5} \and Shahram Abbassi \inst{1,}\inst{6}
          }

   \institute{Department of Physics, Faculty of Science, Ferdowsi University of Mashhad, P.O. Box 1436, Mashhad, Iran \and School of Astronomy, Institute for Research in Fundamental Sciences (IPM), 19395-5531, Tehran, Iran 
              \and Instituto de Astrofísica de Canarias, Calle Vía Láctea s/n, E-38205 La Laguna, Tenerife, Spain \and Departamento de Astrofísica de la Universidad de La Laguna, Av Astrofisíco Francisco Sánchez s/n, 38205 La Laguna, Tenerife, Spain \and Departamento de Astronomía, Universidad de La Serena, Av. Raúl Bitrán 1305, La Serena, Chile
 \and Department of Physics and Astronomy, University of Western Ontario, London, Ontario N6A 3K7, Canada
             }
             
%{mroshan@um.ac.ir; abbassi@um.ac.ir,m.rad@ipm.ir; habib@ipm.ir,jalfonso.aguerri@iac.es,virginia.cuomo@uda.cl}
  % Instituto de Astronomía y Ciencias Planetarias, Universidad de Atacama, Avenida Copayapu 485, 1350000 Copiapó, Chile
   \date{\today}

  \abstract
  % context heading (optional)
  % {} leave it empty if necessary  
   {In this paper, the redshift evolution of the galactic bar properties, like the bar length, pattern speed, and bar fraction, has been investigated for simulated galaxies at stellar masses $M_*>10^{10}\, M_{\odot}$ in the cosmological
magnetohydrodynamical simulation TNG50. We focus on the redshift evolution of the bar pattern speeds and \textit{the fast bar tension}. We show that the median value of the pattern speed of the bars increases as the redshift grows. On the other hand, although the median value of the bar length increases over time, the ratio between the corotation radius and the bar radius, namely the $\mathcal{R}=R_{\text{CR}}/R_{\text{bar}}$ parameter, increases as well. In other words, the corotation radius increases with a higher rate compared to the bar length. This directly means that galactic bars slow down with time, or equivalently as the redshift declines. We discuss the possible mechanisms that reduce the pattern speeds in TNG50. We demonstrate that while mergers can have a significant impact on a galaxy's pattern speed, they do not play a crucial role in the overall evolution of mean pattern speed within the redshift range $z\leq 1.0$. Furthermore, we show that the $\mathcal{R}$ parameter does not correlate with the gas fraction. Consequently, the existence of gas in TNG50 does not alleviate the fast bar tension. We show that the mean value of the pattern speed, computed for all the galaxies irrespective of their mass, at $z=1.0$ is $\Omega_p=70.98\pm 2.34$ km s$^{-1}$ kpc$^{-1}$ and reduces to $\Omega_p=33.65 \pm 1.07$ km s$^{-1}$ kpc$^{-1}$ at $z=0.0$. This is a direct prediction by TNG50 that bars at $z=1.0$ rotate faster by a factor of $\sim 2$ compared to bars at $z=0.0$.}
  % aims heading (mandatory)
 %  {}
  % methods heading (mandatory)
 %  {}
  % results heading (mandatory)
 %  {}
  % {}


   \maketitle
%
%-------------------------------------------------------------------

\section{Introduction}
The appearance of galactic stellar bars is directly related to the collective motion of stars on $x_1$ orbits in the inner parts of the galactic disk \citep{con1980,at1992,2014RvMP...86....1S}. This could happen during the bar instability where the propagation of density waves within the surface of the disk undergoes swing amplification. It is well-understood that the formation and evolution of the bars are directly influenced by the dark matter halo. Whereas the dark matter halo, in the first place, was introduced to suppress bar instability \citep{1973ApJ...186..467O}, it is still true that the live dark matter halo stabilizes the submaximal disks and prevents bar formation by reducing the relative strength of the disk's self-gravity. Recently, in \cite{2022MNRAS.tmpL.138K} it has been shown that even in the cosmological hydrodynamic simulation TNG50 where several baryonic feedbacks are included, the submaximal subhalos suppress the bar formation. It seems to conflict with the barred galaxies in the SPARC database \citep{2016AJ....152..157L}. Interestingly, in maximal or less submaximal disks the situation is completely different, and the existence of a spherical live halo can even trigger the bar instability by exchanging angular momentum with the disk \citep{2002ApJ...569L..83A}. Anyway,  the bar formation is a complex galactic process that depends not only on the dark matter halo but also on the gas fraction of the disk, how hot/cold is the disk \citep{atan2013}.

The pattern speed is affected by the dark matter halo as well. The dynamical friction caused by the dark matter particles slows down the pattern speed of the bar. This fact has been reported in both theoretical investigations \citep{1985MNRAS.213..451W} and N-body galactic simulations \citep{Debattista2000}. On the other hand, cosmological simulations are the best place to investigate this issue in the sense that they provide a more realistic situation for galaxy formation and evolution. On the other hand, since there are usually many galaxies formed within these simulations, a more reliable statistical description can be provided. The majority of bars in EAGLE and Illustris cosmological simulations are ``slow'' \citep{Algorry2017,Peschken2019}. {By a slow bar, we conventionally mean that the ratio between the corotation radius and the bar radius, namely the $\mathcal{R}=R_{\text{CR}}/R_{\text{bar}}$ parameter, is larger than 1.4 ($\mathcal{R}>1.4$). Similarly, ``fast'' bars are defined as bars with $\mathcal{R}<1.4$.}

In addition, the direct comparison between cosmological simulations like EAGLE and IllustrisTNG and the bar pattern speed observations has quantified a serious tension for the standard cosmological $\Lambda$CDM model \citep{2021MNRAS.508..926R}. To be specific, galactic bars at redshift $z=0.0$ in cosmological simulations EAGLE and IllustrisTNG are mostly slow. Whereas almost all the observed galactic bars are fast \citep{Cuomo2020}. The tension exceeds $5\sigma$. Albeit it is necessary to mention that the pattern speed of 225 barred galaxies has recently been measured by \cite{zoo}. {This study utilizes a larger sample of galaxies compared to \cite{Cuomo2020}, which may help mitigate some biases. They found that 62\% of the bars in their sample are slow. However, the mean value of the $\mathcal{R}$ parameter is $\bar{\mathcal{R}} \approx 1.7$, which still differs significantly from that of TNG50 at $z=0$, where $\bar{\mathcal{R}} \approx 3.1$. It is crucial to note that the methods employed by \cite{zoo} to measure bar properties differ from those used by \cite{Cuomo2020}. As a result, caution must be exercised when comparing the findings of these two papers.} 

It is interesting to mention that although there are differences in physics implementations in FIRE2 (Feedback in Realistic Environments) cosmological simulation compared to EAGLE and IllustrisTNG, the same tension appears in FIRE2 as well. In \cite{fire2} the pattern speed of the 13 high resolution Milky way mass galaxies from the zoomed-in simulation FIRE2 has been studied. There are six galaxies with well measured $\mathcal{R}$ parameter. They are all slow in the sense that $\mathcal{R}>1.4$. Although the fact that bars are slow in large-box $\Lambda$CDM cosmological simulations remains challenging, the situation with the Auriga zoom-in simulation \citep{Grand2017} seems to be different. Using this simulation, \cite{Fragkoudi2021} finds bars that remain fast. However, the cost is that bars should grow in galaxies that have higher stellar-to-dark matter ratios. In other words, host galaxies violate the commonly used abundance matching relation in the sense that they are more baryonic-dominated systems.

The fast bar tension may put strict constraints on the properties of dark matter particles. For example, to alleviate this tension, it might be better to postulate dark matter particles that do not lead to a substantial amount of dynamical friction. From this perspective, the ultra-light axion seems to be a good candidate \citep{Hui_2016}. However, there are other observational problems with this candidate. See \cite{Rogers_2021} for further details.  
  \begin{figure*}
  \centering
  \includegraphics[width=0.3\textwidth]{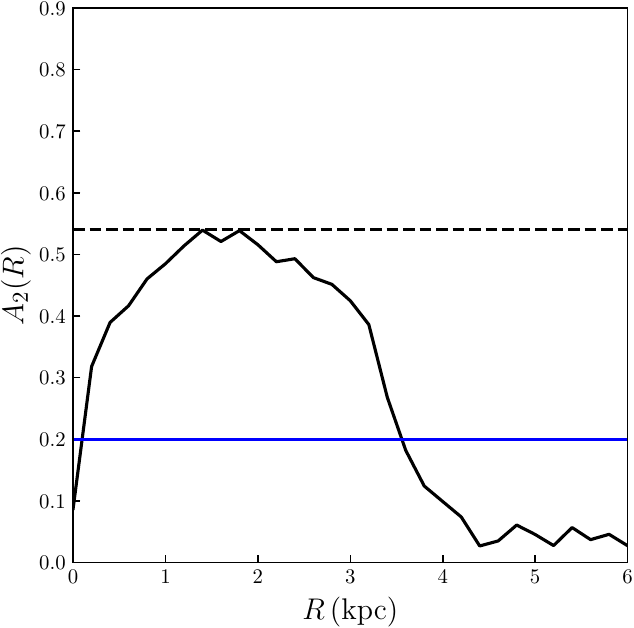}
  \includegraphics[width=0.31\textwidth]{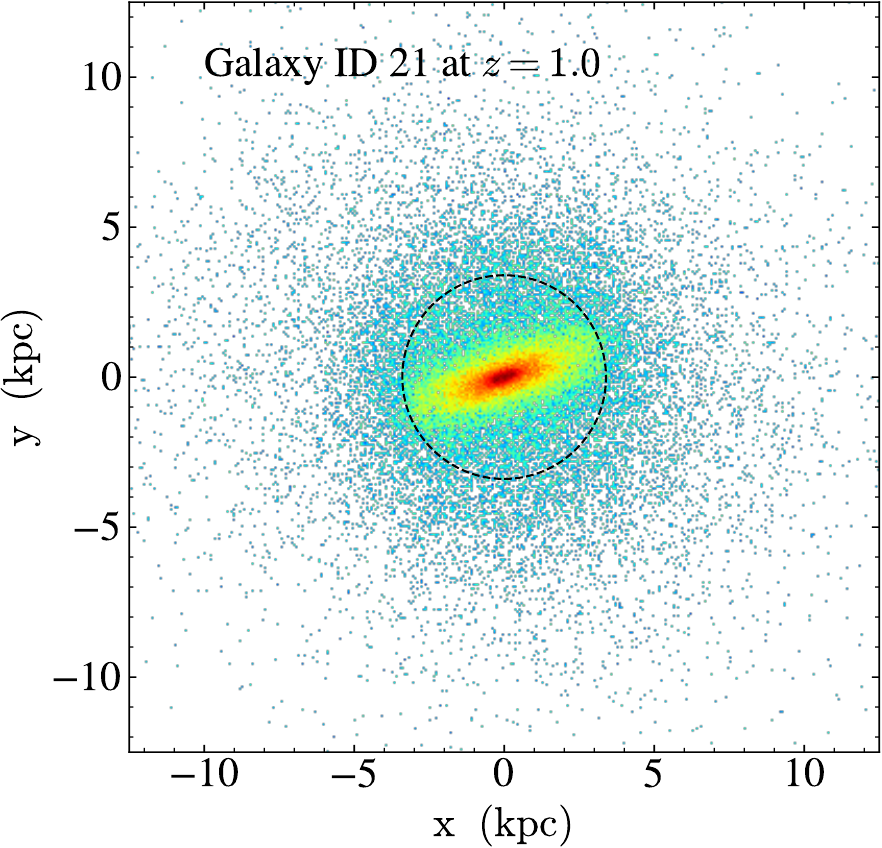}
  \includegraphics[width=0.3\textwidth]{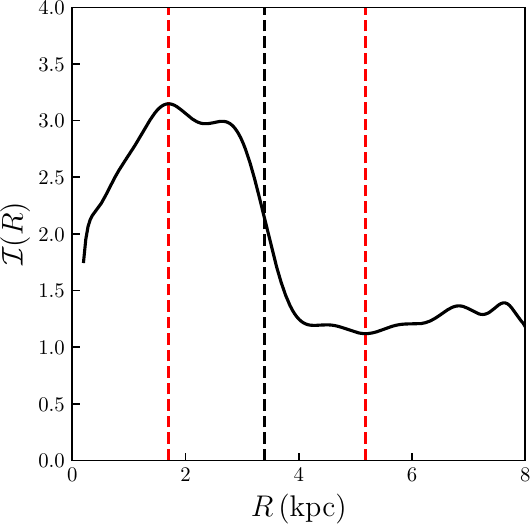}\\
  \includegraphics[width=0.3\textwidth]{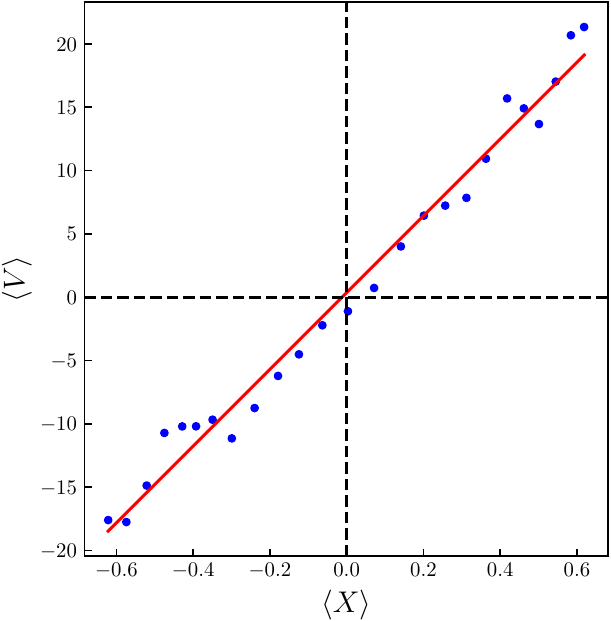}
  \includegraphics[width=0.31\textwidth]{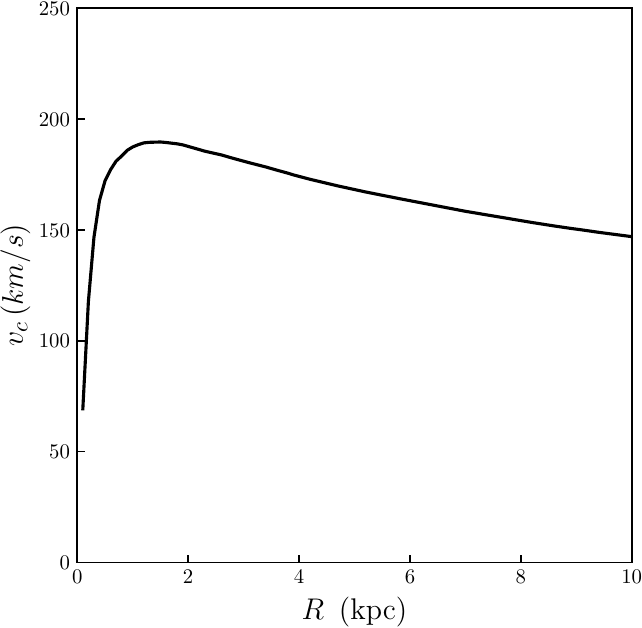}
  \includegraphics[width=0.31\textwidth]{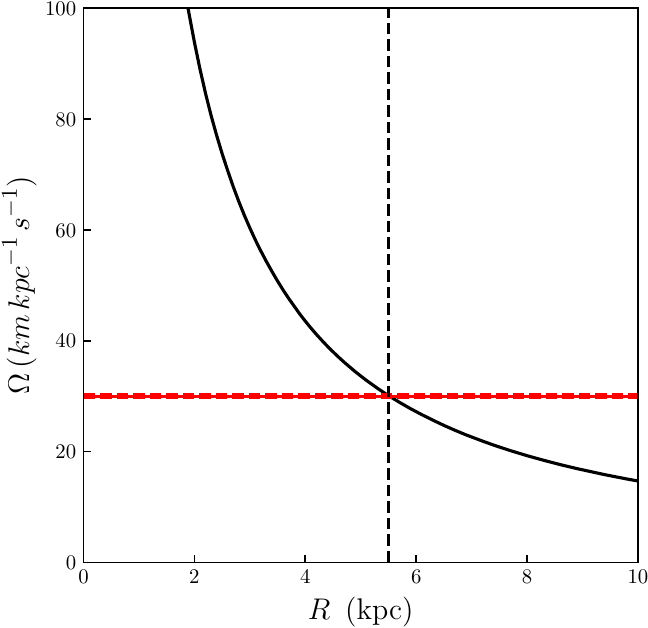}
  \caption{Properties of Galaxy ID 21 in TNG50 at $z=1.0$ as a representative. The upper left panel shows $A_{2}(R)$. The dashed horizontal line indicates $A_{2}^{\text{max}}$. The blue horizontal line $A_{2}=0.2$ indicates the border between unbarred and barred galaxies. The upper middle panel is the projected face-on view of the galaxy. The dashed circle indicates the bar radius obtained by the Fourier analysis. In the right panel, the characteristics of the Fourier method to measure the bar length are illustrated. The vertical red lines show the position of the minimum and maximum values of $\mathcal{I}(R)$. On the other hand, the black dashed line indicates the semi-major axis of the bar. The lower left panel illustrates the TW method when $N_s=25$. The slope gives the pattern speed. The lower middle panel is the total rotation curve. The right panel is the angular velocity. The red dashed line is the pattern speed of the bar and the vertical dashed line indicates the location of the corotation radius.  }
 \label{id21}
\end{figure*}

The dynamical friction has been introduced as the origin of the discrepancy \citep{2021MNRAS.508..926R}. If so, then the pattern speed of the galaxies must increase as redshift grows. There is another possibility that the bars are slow at higher redshifts as well and remain slow until the present time. In this case, the dynamical friction originating from dark matter particles is not responsible for the slowness of the bars at $z=0.0$. To better understand this issue, it is necessary to study the simulated galactic bars at higher redshifts. This is the main purpose of this paper. To do so, we study barred galaxies at stellar masses $M_*>10^{10.0}\, M_{\odot}$ in TNG50 \citep{TNG501,TNG502,TNG503} at redshifts $z=0, 0.5$ and $1$. We restrict ourselves to $z\leq 1$ because the median bar formation redshift is $\lesssim 1$ in TNG50 \citep{Villalba2022}. It might be useful to mention that the age of the Universe at $z=1.0$ ($z=0.5$) is $\approx 5.9$ Gyr ($\approx 8.6$ Gyr). From observational standpoint, it should be mentioned that the only attempt so far to measure the pattern speed as a function of time is \cite{Perez2012}. In this paper, the stellar rings are used as tracers of resonances. Doing this the bars were compatible to be fast up to redshift about 0.5. 

The outline of the paper is as follows: in Section \ref{methods}, we first discuss our galaxy sample selection rules. Then we briefly discuss all the methods used to measure the bar length, strength, pattern speed, and corotation radius. In Section \ref{results}, we present the results and discuss the redshift evolution of the pattern speed. Furthermore, we discuss the possible mechanisms that slow down the bars. Finally, our conclusions are presented in Section \ref{conclusion}.

\section{Methods}
\label{methods}
In this section, we describe the methods implemented to select the galaxy sample, to measure the bar length, strength, and pattern speed. The methods are similar to those in \cite{2021MNRAS.508..926R} where we refer the reader to more details and references. {It is important to note that we employ techniques commonly used in observational studies to measure the characteristics of the bar.} Let's briefly discuss the methods.

\subsection{Selecting the galaxy sample}
\label{sample-selection}
Our sample includes galaxies with stellar mass $M_*> 10^{10}\, M_\odot$. To select the barred galactic disks, we need to use an explicit definition for the disk. First, we compute the direction of the total angular momentum vector for the stellar particles within the stellar half-mass radius. We set the $z$-axis along this direction, and apply two additional widely used criteria to select the disks: i) $k_{\rm{rot}}\geq 0.5$ and ii) $F\leq 0.7$, where $k_{\rm{rot}}$ is a measure of the fraction of rotational kinetic energy to total kinetic energy. To be specific, it is defined as the mass-weighted mean value of $v_{\phi}^2/v^2$ within $30\,$kpc, where $v$ is the total velocity and $v_{\phi}$ is the azimuthal velocity for each stellar particle. On the other hand, the morphological flatness parameter $F$ is defined as $F \equiv M_1/\sqrt{M_2\, M_3}$, where $M_i$s are the eigenvalues of the moment of inertia tensor sorted as $M_1 \leq M_2 \leq M_3$.

\subsection{Bar strength measurement}
\label{barred_disks}
To identify the barred galaxies, we compute the $\rm{m} = 2$ azimuthal Fourier component of the mass distribution of the disk. We consider all the disk particles with $\left| z \right| < 1$ kpc and project all of them to the $x-y$ plane. We divide the projected disk into annuli of fixed width $\delta r=0.1$ kpc and compute the following Fourier coefficients
\begin{eqnarray}
	a_{\rm{m}} \left( R \right) &\equiv& \frac{1}{M \left( R \right)} \sum_{k=0}^{N} m_k \cos \left( \rm{m} \phi_k \right), ~ \rm{m} = 1, 2, .. \, , \\
	b_{\rm{m}} \left( R \right) &\equiv& \frac{1}{M \left( R \right)} \sum_{k=0}^{N} m_k \sin \left( \rm{m} \phi_k \right), ~ \rm{m} = 1, 2, .. \, , 
\end{eqnarray}
where $N$ is the number of particles in the annulus, $R$ is the mean cylindrical radius of the annulus, and $M$ is the total mass of the particles inside the annulus. Each particle is labeled by the index $k$ with mass $m_k$ and azimuthal angle $\phi_k$. In this way, the Fourier amplitude for mode $\rm{m}$ at radius $R$ is defined as
\begin{eqnarray}
    A_{\rm{m}} \left( R \right) ~\equiv~ \sqrt{a_{\rm{m}} \left( R \right)^2 + b_{\rm{m}} \left( R \right)^2} \, .
\end{eqnarray}
On the other hand, $A_0=0.5$ \citep{ohta,Aguerri_2000}. When an $\rm{m}=2$ symmetric feature like bar exists in the disk, this function normally has an evident maximum, see the upper left panel in Fig. \ref{id21} for Galaxy ID 21 in TNG50 at $z=1.0$. This maximum value is taken to define the bar strength
\begin{eqnarray}
    A_2^{\text{max}} ~\equiv~ \max [A_2(R)] \, .
    \label{A2_max}
\end{eqnarray}
Bars are commonly divided into two categories: strong bars with $A_2^{\text{max}} \geq 0.4$, and weak bars with $0.2 \leq A_2^{\text{max}} < 0.4$. On the other hand, disks with $A_2^{\text{max}} < 0.2$ are unbarred.

\subsection{Bar length measurement}
\label{bar-ln}

To measure the bar length we use the Fourier decomposition of the surface density \citep{Aguerri_2000}. We first compute the intensity in the bar ($I_b$) and inter-bar ($I_{ib}$) zones, and find the following ratio as a function of radius $R$:
\begin{equation}
	\mathcal{I} \left( R \right) ~\equiv~ \frac{I_b \left( R \right)}{I_{ib} \left( R \right)} ~=~ \frac{A_0 ~+~ A_2 ~+~ A_4 ~+~ A_6}{A_0 ~-~ A_2 ~+~ A_4 ~-~ A_6 } \, .
	\label{Bar_interbar_ratio}
\end{equation}
Finally, the semi-major axis of the bar is the outer radius beyond which $\mathcal{I} \left( R \right) < \left( \mathcal{I}^{\text{max}} + \mathcal{I}^{\text{min}} \right)/2$, where $\mathcal{I}^{\text{max}}$ and $\mathcal{I}^{\text{min}}$ are the maximum and minimum values of $\mathcal{I} \left( R \right)$ respectively. The bar length is twice the semi-major axis, see the upper right panel in Fig. \ref{id21}. We take $\delta r=0.1\,$kpc as the error of the bar length.
\subsection{Pattern speed measurement}
\label{TW-method}

In the isolated simulations where the full-time evolution of the disk is accessible, one can easily measure the pattern speed by finding the bar position angle in terms of time. However, here we have a single snapshot of the galaxies. Therefore, we need to treat them like real galaxies where the Tremaine-Weinberg method (TW) \citep{Tremaine1984} is widely used to measure the pattern speed. To use this method in simulations, we need the surface density $\Sigma$, the line of sight velocity $V_{\text{LOS}}$ and the positions of the particles/stars, and the inclination angle $i$ of the galaxy. The pattern speed $\Omega_p$ is then $\Omega_p \sin i = \langle V \rangle/\langle X \rangle$, where 
\begin{eqnarray}
    \langle V \rangle ~&\equiv&~ \frac{\int V_{\text{LOS}} \Sigma \, dX}{\int \Sigma \, dX} \, , \\
    \langle X \rangle ~&\equiv&~ \frac{\int X \Sigma \, dX}{\int \Sigma \, dX}
\end{eqnarray}
 $\langle V \rangle$ and $\langle X \rangle$  are the so-called kinematic and photometric integrals, defined as the luminosity-weighted average LOS velocity $V_{\rm LOS}$ and position $X$ parallel to the major axis of the particles/stars, respectively. The integrals are taken along slits parallel to the disk's major axis. We take $N_s$ evenly spaced slits with length $l_s$ and width $\Delta_s$. Slits are distributed only in the bar region specified with the height $h_s$. More specifically, $h_s$ is identical to the projection of the bar length along the minor axis of the disk. We start with the face-on view of the galaxy and set the position angle of the bar at $60^\circ$ relative to the major axis of the disk. Then we rotate the galaxy and fix it at inclination angle $i = 45^\circ$. Finally, we start with a fixed set of $(N_s, \Delta_s,h_s)$, and vary the slit length $l_s$ until the pattern speed converges to a unique value. 

Linear regression on $\langle V \rangle$ and $\langle X \rangle$ is used to find the slope $\Omega_p$; see the lower left panel in Fig. \ref{id21}. The standard error of the regression is defined as the error on $\Omega_p$. We then repeat the process taking different sets of $(N_s, \Delta_s,h_s)$. In this way, we find several estimates for the pattern speed. Finally, the mean value of the pattern speed is our final $\Omega_p$. On the other hand, the largest deviation from the mean is taken as the error of $\Omega_p$. In some cases, the error exceeds $20$\%. We remove those cases to achieve reliable values for the pattern speeds. The reliability of our TW code has been already tested in \cite{2021MNRAS.508..926R}. To be specific, a galaxy was simulated using the GALAXY code \citep{sellwood_code} where the pattern speed evolution is accurately known. Then our TW code has been used to compute the time evolution of the pattern speed. A full comparison has been presented in \cite{2021MNRAS.508..926R} implying that the code works reliably. 

It is worth noting that a more accurate method for measuring pattern speeds in simulations has recently been introduced in \cite{newps}. This method utilizes all the position and velocity information of the particles, not just the line of sight velocity. Applying this new method to measure pattern speeds in cosmological simulations would be the subject of another study. Here, we adhere to the TW method since we treat simulated galaxies as real galaxies, for which we only have the line of sight velocity.

\subsection{Corotation radius measurement}
\label{CR}
To find the corotation radius $R_{\text{CR}}$ where $\Omega_p=\Omega(R_{\text{CR}})$, we need the rotation curve $v_c(R)$ of the galaxy. $\Omega(R)$ is the angular velocity of the disk defined as $v_{c}(R)=R\Omega(R)$. One of the widely used methods to obtain the rotation curve in cosmological simulations is to measure the total mass within the radius $R$, namely $M(R)$. This mass contains all the mass contributions from baryonic matter and dark matter. Then the rotation curve is simply estimated as
\begin{equation}
v_c(R)=\sqrt{\frac{G M(R)}{R}}
\label{eight}
\end{equation}
In this way, the radius at which the bar pattern speed intersects with $v_c/R$, indicates the corotation radius. It should be noted that the error in pattern speed leads to an error in corotation radius. One may use the particle accelerations to find a more accurate circular velocity, $v_c^*$, obtained by computing the particle accelerations. However, equation \eqref{eight} recovers $v_c^*$ at $R>2$ kpc with an accuracy better than 4\% \citep{2021MNRAS.508..926R}. Since most of the galaxies in TNG50 have $R_{\text{CR}}>2$ kpc, $v_c$ given by \eqref{eight} is a suitable choice to locate the corotation radius.

\subsection{Observed sample}
{To perform a consistent comparison with observations in the local Universe, we defined a sample of barred galaxies at $z<0.1$, with a stellar mass range comparable with our simulations, and for which the bar and galaxy properties (bar length, strength and pattern speed and general properties) have been derived with the same approaches adopted here and described in Sec.~\ref{methods}. We adopted the results presented by \citet{Cuomo2020}, where a sample of 77 barred galaxies was analysed. The sample contains 43 late-type barred galaxies and 34 early-type barred galaxies. From their sample, we considered only the subsample of galaxies, included as well in the CALIFA \citep{sanchez2012} and MaNGA \citep{bundy2015} surveys, for which stellar mass estimates are available and provided by \cite{Bitsakis2019} and \cite{Sanchez2022}, for the CALIFA and MaNGA surveys, respectively. Therefore, we refer to the CALIFA+MaNGA sample in the following discussions, which include 60 barred galaxies, with stellar masses $>10^{10}~M_{\odot}$ that satisfy the condition $\Delta\Omega_p/\Omega_p\leq 0.5$. The galaxies of the CALIFA subsample have Hubble types
ranging from SB0 to SBd, and redshifts $0.005 < z < 0.03$. On the other hand, the galaxies in the MaNGA subsample have Hubble types ranging from SB0 to SBc, redshifts $0.02<z<0.1$. In both subsamaples we have weak and strong bars as in TNG50. }

{In the CALIFA and MaNGA surveys, the analysis is primarily influenced by the selection criteria of both observational studies, which are based on galaxy mass and volume. Specifically, lenticular and early-to-intermediate spiral galaxies are targeted, as these types are more suitable for the TW method. However, this selection criteria only excludes very late-type barred galaxies, see Fig. 1 in \citet{Cuomo2020}. Regarding the bulge characteristics, the observational sample encompasses a range of bulge types, from bulgeless galaxies to those with significant bulges (up to $B/T = 0.6$), as illustrated in Fig. 6 of \citet{Cuomo2020}. Thus, the observational sample is quite comprehensive in terms of bulge types. For the identification of disks within the observational sample, the photometric morphological classifications provided by the surveys are used. Additionally, the stellar kinematic two-dimensional maps are utilized to select disks that are rotation-dominated.}

{
In \citet{Cuomo2020}, observed bars were identified through a two-step process. First, the galaxy optical images were visually inspected, followed by Fourier analysis to identify the peak in the $m=2$ Fourier components. The presence of the bar was then confirmed using the TW method, which produced a TW signal. In contrast, \citet{zoo} relied solely on visual inspection of the images to identify bars.}

{We emphasize that we maintain a consistent comparison between observations and simulations by employing the same methodologies to measure galactic bar properties in both contexts. Furthermore, the galaxies in both scenarios fall within the same mass range and encompass a similar distribution of weak and strong bars. However, this approach may introduce some biases. For instance, there may be a potential bias related to the use of the TW method. In actual observations, we cannot achieve the same level of precision as we do in simulations. In TNG50, we obtain accurate measurements of the bar's position angle and the inclination angle of the simulated galaxies. Unfortunately, this degree of precision is often absent in observational data, which can lead to biases in our findings. A more comprehensive approach could involve generating mock observations from TNG50 that replicate MaNGA and CALIFA data, followed by the application of the TW method to these mock observations. However, this falls outside the scope of our current paper.}
\section{results}
\label{results}
%Before discussing the results, it is important to emphasize that while we have attempted to apply techniques used in observations, there are significant differences in their application to simulated and real galaxies. For instance, in simulated galaxies, we were able to accurately measure the disk position and inclination angles, which is not the case in observations, introducing there unavoidable sources of errors.
\begin{table}
	\centering
	\begin{tabular}{ccccc}
		\hline
		\multirow{2}{*}{} & $z=0.0$ & $z=0.5$ & $z=1.0$ \\
		\hline
		%Total & 903 & 740 & 597  \\
		Disks & 609 & 568 & 389  \\ 
		Barred & 258 & 285 & 199  \\
		Strong & 125 & 128 & 79  \\
		Weak & 133 & 157 & 120  \\
		Total bar fraction & 0.42$\pm$ 0.02 & 0.50$\pm$ 0.02 & 0.51$\pm$ 0.02\\ 
		With reliable $\Omega_p$ & 200 & 251 & 167 \\
		\hline
	\end{tabular}
	\caption{Bar statistics in TNG50 at different redshifts based on galaxies with stellar mass $M_*> 10^{10.0} \, M_\odot$. Disks are identified with the selection rules $k_{\rm{rot}}\geq 0.5$ and $F \leq 0.7$ (Section \ref{sample-selection}). The value of $A_2^{\text{max}}$ (Section \ref{barred_disks}) is used to classify galaxies as unbarred ($A_2^{\text{max}}<0.2$), weakly barred ($0.2\leq A_2^{\text{max}} < 0.4$), or strongly barred ($A_2^{\text{max}} \geq 0.4$). The total bar fraction is the ratio of the number of barred disks to the total number of disks. Reliable pattern speeds are those with errors less that 20\%. }
	\label{Sample_sizes}
\end{table}
We summarise the bar statistics in Table \ref{Sample_sizes}. Our selection method shows that the number of disk galaxies grows with time. This is consistent with the results of \cite{Rosas2022}, while different selection methods are used. The number of barred disks increases from $z=1.0$ to $z=0.5$. On the other hand, from $z=0.5$ to $z=0.0$, the number of barred disks decreases with by factor of about $10$\%. This reduction with a higher rate is also reported in \cite{Rosas2022}. Accordingly, the total bar fraction, namely the ratio of the number of barred disks to the total number of the disk galaxies, is almost constant in $0.5<z<1.0$. At this redshift interval, $50$\% of the disks are barred. However, this fraction reduces to $42$\% at $z=0.0$. Anyway, we see that the bar fraction increases with redshift. The same trend has been reported in \cite{Rosas2022} for TNG50.

It should be noted that in isolated simulations there is a common feature almost always seen in maximal disks: after the bar instability, the buckling of the disk reduces the strength of the bar. However, at longer times of the evolution, the interaction between the halo and the disk increases the strength of the bar. Therefore, we do not expect that the bar gradually fades away in the isolated simulations. Consequently, the reduction in the number of barred galaxies in TNG50 must be related to cosmological features like the interaction between galaxies, mass accretion, and implemented baryonic feedback \citep{2024A&A...684A.179R}.

\subsection{Bar fraction in TNG50}
\label{fbar}
\label{bar_rad}
  \begin{figure}
  \centering
  \includegraphics[width=0.45\textwidth]{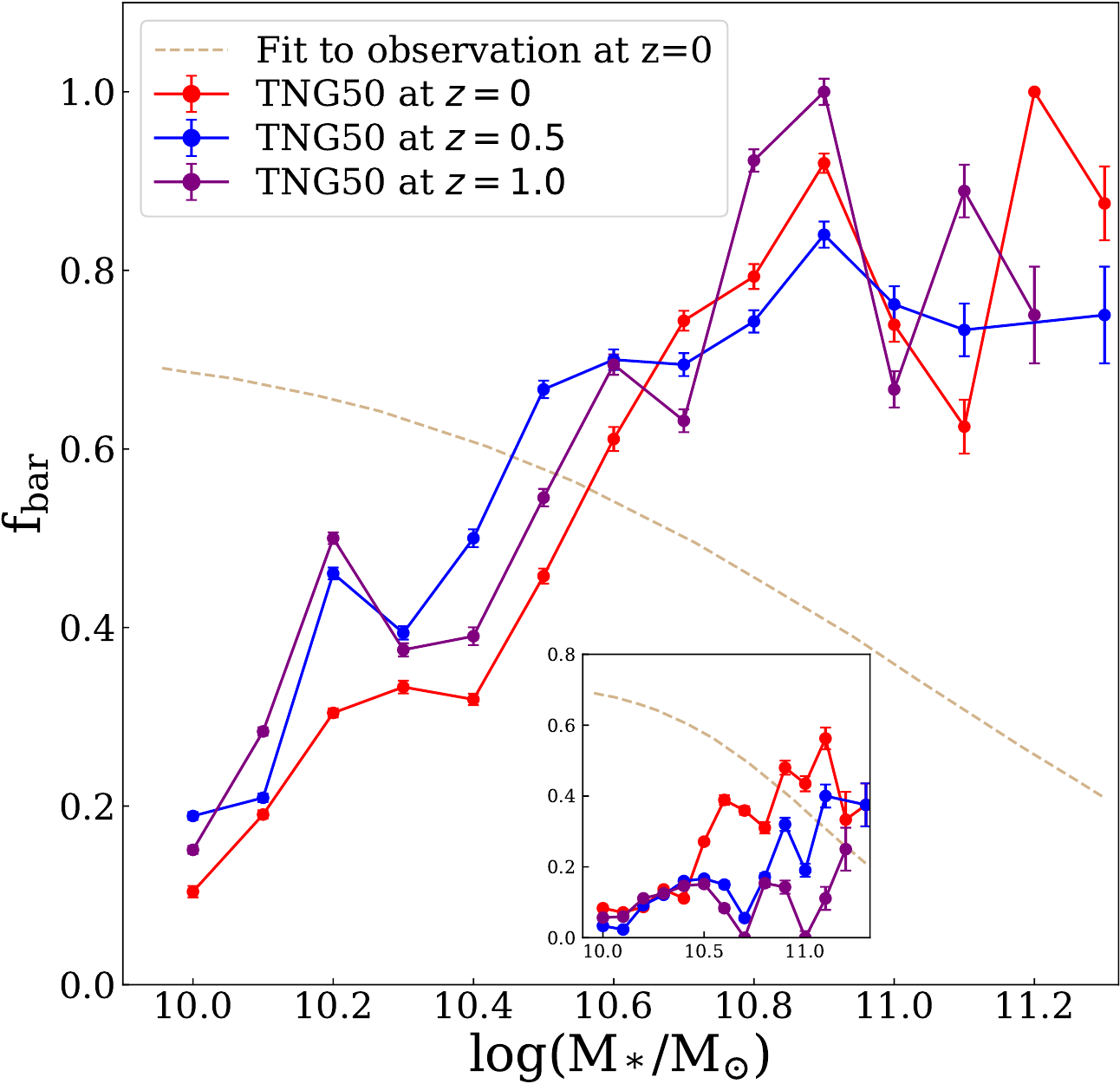}
  \caption{The bar fraction vs stellar mass  at $z=0.0$, $z=0.5$, and $z=1.0$ in TNG50. The dashed curve indicates the fit to observations given in \citet{Erwin2018}. We have taken this curve from \citet{Zhao2020}. The binomials errors on the bar fractions are defined using the bar fraction and the number of total disks, i.e., $n_{\text{disk}}$, in each mass interval bin as $\Delta=\sqrt{\frac{f_{\text{bar}}(1-f_{\text{bar}})}{n_{\text{disk}}}}$ \citep{Rosas2022}. The curve for $z=0.0$ differs slightly from that shown in Fig. 1 of \citet{2021MNRAS.508..926R}, where the condition $F \leq 0.5$ was used instead of $F \leq 0.7$. The inset plot displays the bar fraction for bars longer than 2 kpc. This plot is useful because very short bars may not be detected in observations.}
 \label{fig1}
\end{figure}
The redshift evolution of the bars in galaxies provides information about the time when the rotation dominates the dynamics of the galaxies \citep{Sheth2012}. Moreover, it would be interesting to measure the bar fraction $f_{\text{bar}}$ in terms of the stellar mass of the galaxies. In this way, it would be easier to compare TNG50 with the results of other simulations and with the relevant observations. We performed the bar analysis for
 $z=0.0$, 
 $z=0.5$ 
 and 
 $z=1.0$ 
 for galaxies in the stellar mass range $M_*\gtrsim 10^{10}\, M_{\odot}$. It can be seen in Fig. \ref{fig1} that the bar fraction is strongly correlated with the stellar mass of the galaxy. %At $z=0.0$ the bar fraction peaks around $M_*\simeq  10^{11.2}\, M_{\odot}$, which is consistent with the results reported by \cite{Zana2022}. 
 
On the other hand, for galaxies with masses $M_*\lesssim 10^{10.6}\, M_{\odot}$, the bar fraction dramatically decreases with time.  On the other hand, for galaxies with $M_*\gtrsim 10^{10.6}\, M_{\odot}$, there is no smooth behaviour in terms of redshift. As an example, for $M_* \simeq 10^{10.7}\, M_{\odot}$ the bar fraction reaches from about $60\%$ at $z=1.0$ to about $75\%$ at $z=0.0$.  
 
From  an observational standpoint, all observations agree that the fraction of bar galaxies in our local universe must be high \citep{Eskridge2000,Whyte2000,Lauri2004,Menedez2007,Marinova2007,Barazza2008,Sheth2008,Aguerri2009,Nair2010,Masters2012,Melvin2014,Diaz2016}. Meanwhile, there is always a debate about the dependence of the fraction of barred galaxies on the stellar mass. Most SDSS-based studies suggest that the bar fraction increases strongly for $M_*\gtrsim 10^{10}\, M_{\odot}$ \citep{Masters2012,Oh2012,Melvin2014}. Contrary to these results, \citet{Erwin2018} reports that for a sample of nearby galaxies $(z \lesssim 0.01)$ in $S^{4}G$ survey, the bar fraction reaches a maximum of about $76\%$  for $M_* = 10^{9.7}\, M_{\odot}$ from $20\%$ for galaxies with low stellar mass $(M_* = 10^{8}\, M_{\odot})$  and then in $M_* = 10^{11}\, M_{\odot}$ decreases again to $40\%$. \citet{Erwin2018} explains the inconsistency between the $S^{4}G $ survey results and the SDSS studies by saying that this inconsistency is probably caused by the inefficiency of the SDSS studies in identifying small bars at low stellar masses $\lesssim 10^{10}\, M_{\odot}$.
  
In contrast to the nearby universe, measuring the bar fraction at higher redshifts is more difficult due to the lack of sufficient resolution and band-shifting \citep{Sheth2003}. {The bar length can be significantly altered due to band-shifting \citep{band-shifting}}. But today, thanks to high-resolution observations of the deep visible and near-infrared regions, we know that the fraction of barred galaxies decreases with redshift \citep{Sheth2008,Cameron2010,Melvin2014,Simmons2014}. It should be noted that these studies were carried out at higher redshifts on galaxies with stellar masses higher than $10^{10}\, M_{\odot}$. However, we see in Fig. \ref{fig1} that for a wide mass interval $10^{10}\, M_{\odot} \lesssim M_* \lesssim 10^{10.6}\, M_{\odot}$ the bar fraction increases with redshift.

  \begin{figure}
  \centering
  \includegraphics[width=0.45\textwidth]{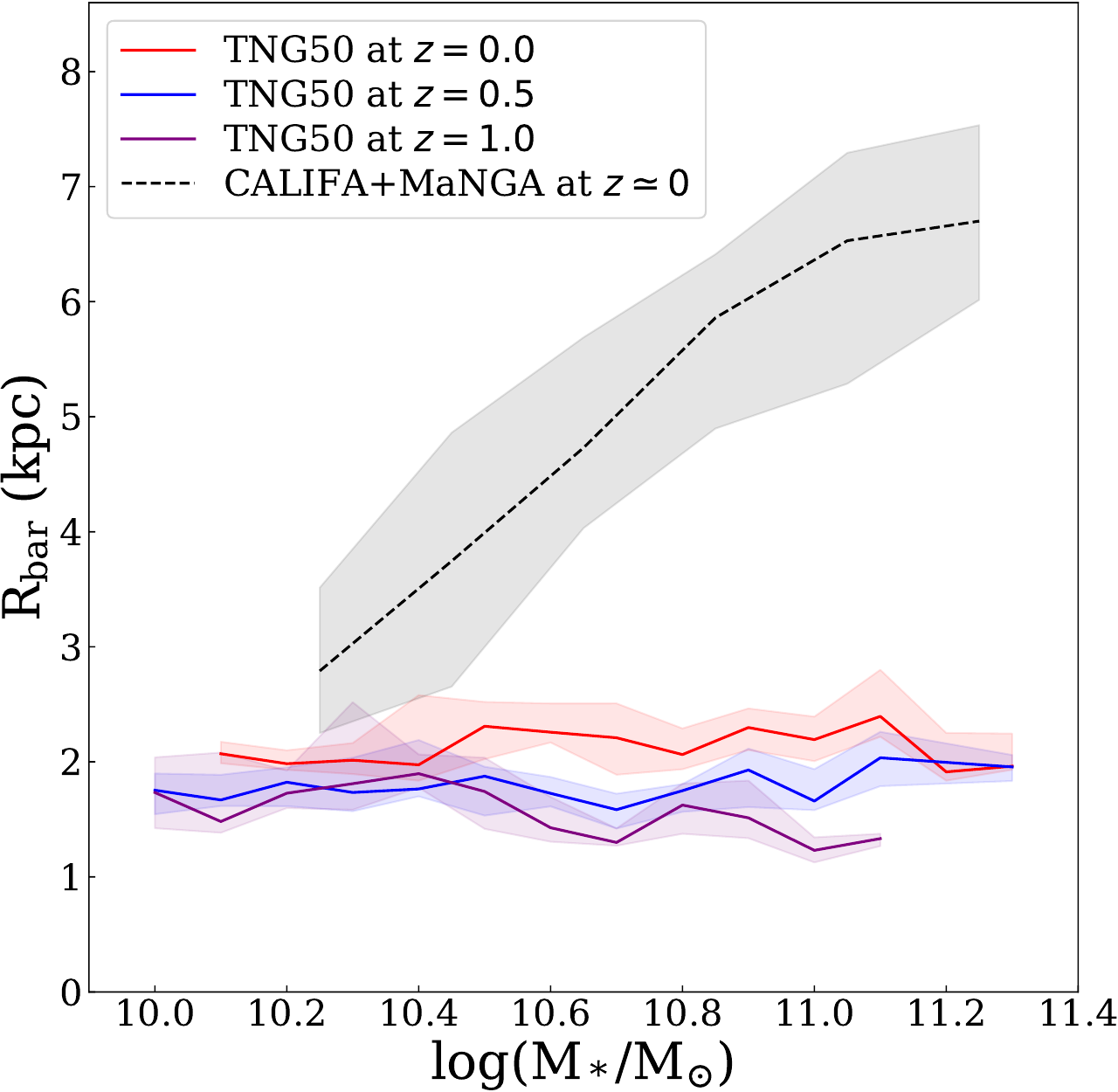}
  \caption{The solid lines display the median value of the bar radius in kpc vs stellar mass for TNG50 at $z=0.0$, $z=0.5$, and $z=1.0$ and for the observed CALIFA+MaNGA sample at $z \simeq 0.0$. We have included 20 additional galaxies from CALIFA and MaNGA that do not meet the condition $\Delta \Omega_p/\Omega_p \leq 0.5$, but satisfy $\Delta R_{\text{bar}}/R_{\text{bar}} \leq 0.5$. The shaded areas give the $32^{\text{nd}}$- $68^{\text{th}}$ percentile. We only display mass bins that include more than five galaxies. We plot the same regions for other similar figures in this paper.}
 \label{l_m}
\end{figure}

It is necessary to mention that our results are in relatively good agreement with that of \citet{Rosas2020}, \citet{Zhou2020}, \citet{Zhao2020}, \citet{Rosas2022} and \citet{Reddish2022}. In the first three works, the bar evolution is investigated in TNG100 simulation. The results by \citet{Rosas2020} show that the bar fraction at $z=0.0$ with a mass range $10^{10.4-11}\, M_{\odot}$ increases with increasing stellar mass. Also, \citet{Zhou2020} made a comparison between the evolution of the bars in Illustris and TNG100 simulations. They found that in both of these simulations, the bar fraction increases with stellar mass for $M_*\gtrsim 10^{10.5}\, M_{\odot}$. In the same line, \citet{Zhao2020} investigated TNG100 galaxies with stellar masses $M_*\gtrsim 10^{10}\, M_{\odot}$, and found the bar fraction increases with stellar mass.
\begin{table}
	\centering
	\begin{tabular}{ccccc}
		\hline
		\multirow{2}{*}{} & $z=0.0$ & $z=0.5$ & $z=1.0$ \\
		\hline
		%Total & 903 & 740 & 597  \\
		$\bar{R}_{\text{bar}}$ (kpc) & 2.30$\pm$ 0.11 & 1.87 $\pm$ 0.14 & 1.71 $\pm$ 0.12  \\
		$\bar{\Omega}_p$  $\left(\frac{\text{km}}{\text{kpc} \, \text{s}}\right)$ & 33.65 $\pm$ 1.07 & 48.61$\pm$1.85 & 70.98$\pm$2.34  \\
		$\bar{\mathcal{R}}$ & $3.10^{+0.33}_{-0.31}$ & $2.82^{+0.36}_{-0.35}$ & $2.31^{+0.34}_{-0.33}$  \\
		$f_{\text{slow}}$&0.93&0.95&0.84\\
		$f_{\text{fast}}$&0.05&0.04&0.13\\
		$f_{\text{ultra-fast}}$&0.02&0.01&0.03\\
	
		\hline
	\end{tabular}
	\caption{The mean value of the bar radius, bar pattern speed, and the $\mathcal{R}$ parameter in terms of redshift. $f_{\text{slow}}$ is the fraction of the slow bars defined by $\mathcal{R}>1.4$ , $f_{\text{fast}}$ is the fraction of fast bars defined by $1<\mathcal{R}<1.4$, and $f_{\text{ultra-fast}}$ is the fraction of ultra-fast bars defined by $\mathcal{R}<1$. }
	\label{mean_values}
\end{table}

All these simulations seem to be in conflict with observations reported in \citet{Erwin2018}.  \citet{Zhou2020} argued that this discrepancy between simulations and observations is due to the low resolution of TNG100 and its inability to identify bars with smaller radii, and also that observation has difficulty in identifying such short bars in high redshifts. However, we see that the same disagreement exists in TNG50, where the resolution is much higher. 

To conclude this section, let us mention that the bar fraction discrepancy appears to exist in all the cosmological simulations. In the case of EAGLE simulations see \citet{Roshan2021}, and for NewHorizon simulation see \citet{Reddish2022}. It is important to note that the discrepancies between the bar fractions observed and those simulated may be attributed to the manner in which the bars are chosen. In observations, bars are selected from non-axisymmetric characteristics in images of disk galaxies. Conversely, in simulations, the selection process is based on dynamical features that take into consideration the galaxy's rotation. Thus, it is possible that the simulations may be selecting galaxies that are not detectable in observations. As previously mentioned, the resolution of the SDSS is not sufficient to resolve bars in low-mass galaxies. Therefore, the observational bar fraction is underestimated for such galaxies. This limitation also applies to other surveys. However, recent work with the James Webb Space Telescope (JWST) has shown that bars longer than 1.3 kpc at $z\sim 1-3$ can be identified by JWST \citep{mobasher}
. This implies that future observations by JWST would give significantly higher bar fractions than previous Hubble Space Telescope (HST) studies. Investigating this possibility is beyond the scope of this paper. However, let us conduct a quick test by removing very short bars ($R_{\text{bar}}<2$ kpc) that are difficult to classify as bars in observations, and then recalculate the bar fraction. The resulting plot is shown in the inset of Fig. \ref{fig1}. As expected, the bar fraction decreases for all stellar masses and approaches the observations at $z=0.0$, at least for massive galaxies. Furthermore, the bar fraction clearly decreases with $z$ for stellar masses greater than $\approx 10^{10.5} M_{\odot}$.

\subsection{Bar radius in TNG50}

We have measured the bar radius in terms of stellar mass and illustrated it in Fig. \ref{l_m}. The solid lines display the median value and the shaded areas give the $32^{\text{nd}}$- $68^{\text{th}}$ percentile. As we can see, the bar length increases towards $z=0.0$. This is also shown in Table \ref{mean_values}, where the mean value of the bar radius is given in terms of redshift {(the mean bar length increases by $\sim 30\%$ from $z=1.0$ to $z=0.0$).} This is quite consistent with the bar length measurement for TNG50 reported in \cite{Rosas2022}, where the median bar length grow from $\approx 1.8$ kpc at $z=1.0$ to $\approx 2.6$ kpc at $z=0.0$. From Fig. \ref{l_m} we see that the bars are {slightly} longer in less massive galaxies at redshift $z=1.0$. However, for other redshifts, there is no tangible correlation between the bar length and stellar mass. Therefore there are two main features in our bar length measurement: i) the mean bar length increases towards $z=0.0$. This is somehow consistent with the results of \cite{Zhao2020} where barred galaxies of cosmological simulation TNG100 were investigated. They found that the bar sizes have grown from $\sim 2\,$kpc at $z=1.0$ to $\sim 3\,$kpc at $z=0.0$. ii) There is no clear correlation between the bar length and {the galaxy} stellar mass. 

Both features seem to conflict with observations. Figure~\ref{l_m} shows the observed bar lengths with respect to the galaxy stellar mass. At $z=0.0$, a strong correlation between these two properties is observed. Moreover, observed bars are much larger than the simulated bars studied here at $z=0.0$, regardless of the stellar mass.
  \begin{figure}
  \centering
  \includegraphics[width=0.47\textwidth]{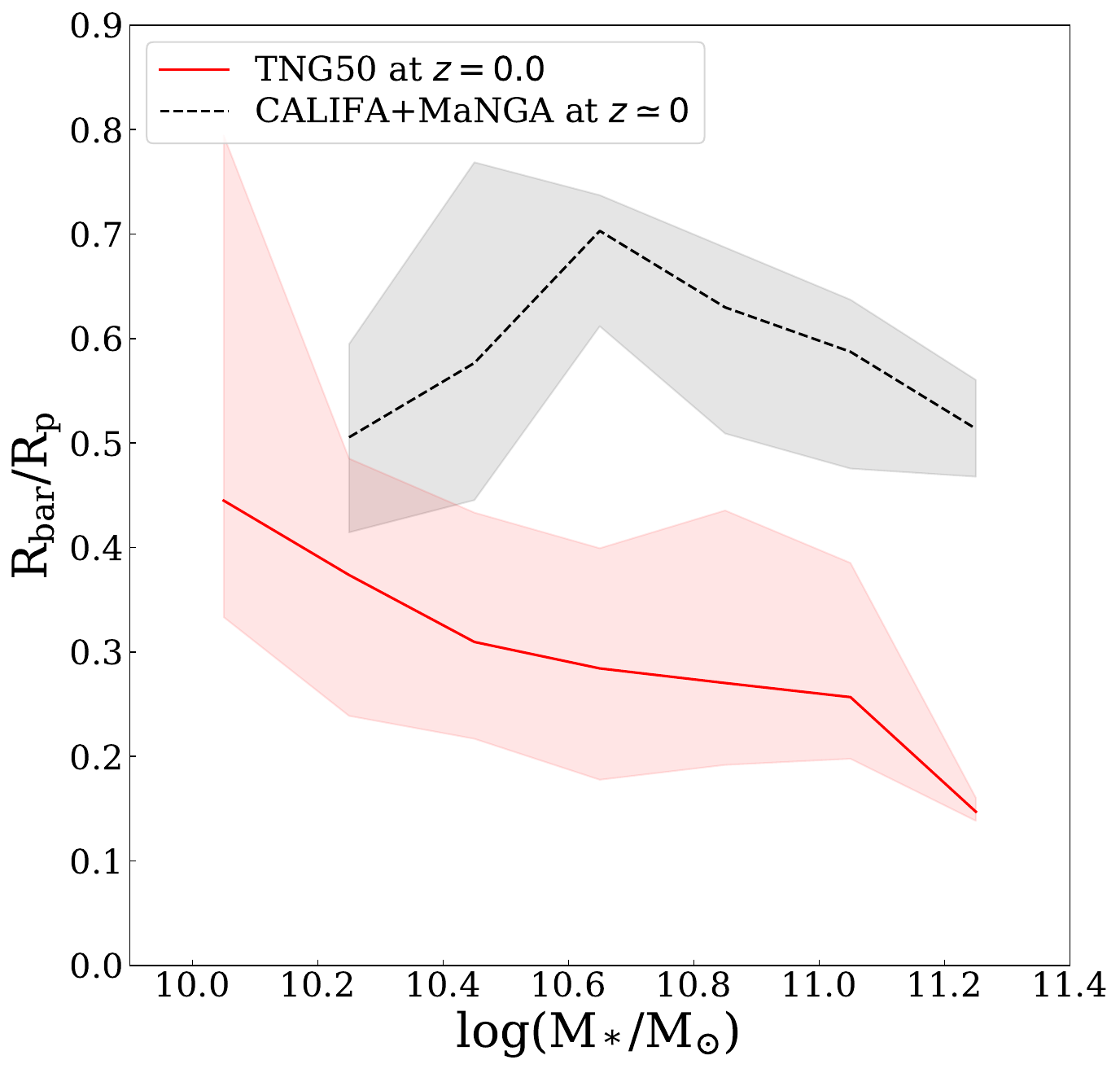}
  \caption{The solid lines display the median value of the scaled bar radius, $R_{\text{bar}}/R_{\text{p}}$, vs stellar mass for TNG50 at $z=0.0$ and for the observed CALIFA+MaNGA sample at $z \simeq 0.0$. $R_{\text{p}}$ is the Petrosian radius used as an indicator for the radial size of the galaxies.}
 \label{scaled_rb}
\end{figure}

{This problem with bar sizes in TNG50 was reported as well by \cite{Frankel2022}.} This paper compared the distribution of bar size in TNG50 at $z=0.0$ to those from MaNGA observations. \cite{Frankel2022} found that the galactic bars in TNG50 are on average 35\% shorter compared to MaNGA observations. The mean value of the bar radius obtained by \cite{Frankel2022} is consistent with our result.

{The comparison with other observational studies reveal further conflicts with the results obtained here. As an example, t}he redshift evolution of 379 barred galaxies at $0.2 < z \leq 0.835$ with $10.0 \leq \log\frac{M_*}{M_{\odot}} \leq 11.4$ from the COSMOS survey has been investigated in \cite{Kim2021}. The observed bar lengths and strengths do not show any clear trend with $z$ and remain roughly constant for the whole redshift interval. On the other hand, the observed bar lengths are strongly correlated with the stellar mass of the host galaxy. Similarly, \cite{Erwin2019} found that although for less massive galaxies ($\log\frac{M_*}{M_{\odot}} \leq 10.1$) the bar length is almost independent of the stellar mass, for massive galaxies ($\log\frac{M_*}{M_{\odot}} > 10.1$), the bar length {strongly depends on} the stellar mass. 

In conclusion, the scaled bar radius is depicted in Fig. \ref{scaled_rb} for the case of observations and TNG50 at $z=0.0$. We utilize the Petrosian radius, $R_\text{p}$, as a reference for the radial size of galaxies in both simulations and observations, scaling the bar length relative to the Petrosian radius. The Petrosian radius is the radius at which the intensity equals $\eta$ times the average intensity. $R_{\text{p}}$ is provided by the SDSS in the r-band, and has been collected for the CALIFA+MaNGA sample in \cite{Cuomo2020}. In TNG50, we find $R_p$ using the following relation
\begin{equation}
\Sigma(R_{\text{p}})=\eta \frac{M(R_{\text{p}})}{\pi R_{\text{p}}^2}
\end{equation}
Here, $\Sigma(R)$ represents the stellar surface density derived from projecting the disk stellar particles onto the disk plane, $M(R)$ denotes the stellar mass enclosed within the radius $R$, and $\eta$ is set to $0.2$. As one may anticipate, Fig. \ref{scaled_rb} illustrates that this scaled bar radius is greater in observations than in TNG50. {We have not included simulated galaxies at $z=1.0$ and $z=0.5$ in Fig. \ref{scaled_rb} because our observational sample consists of nearby galaxies. However, since disks are smaller in size at higher redshifts, and the bar length does not change significantly over time, we would naturally expect the ratio $R_{\text{bar}}/R_\text{p} $ for high redshifts to be greater than the red curve in Fig. \ref{scaled_rb} in the given mass interval.}

\subsection{Pattern speed $\Omega_p$ in TNG50}

  \begin{figure}
  \centering
  \includegraphics[width=0.45\textwidth]{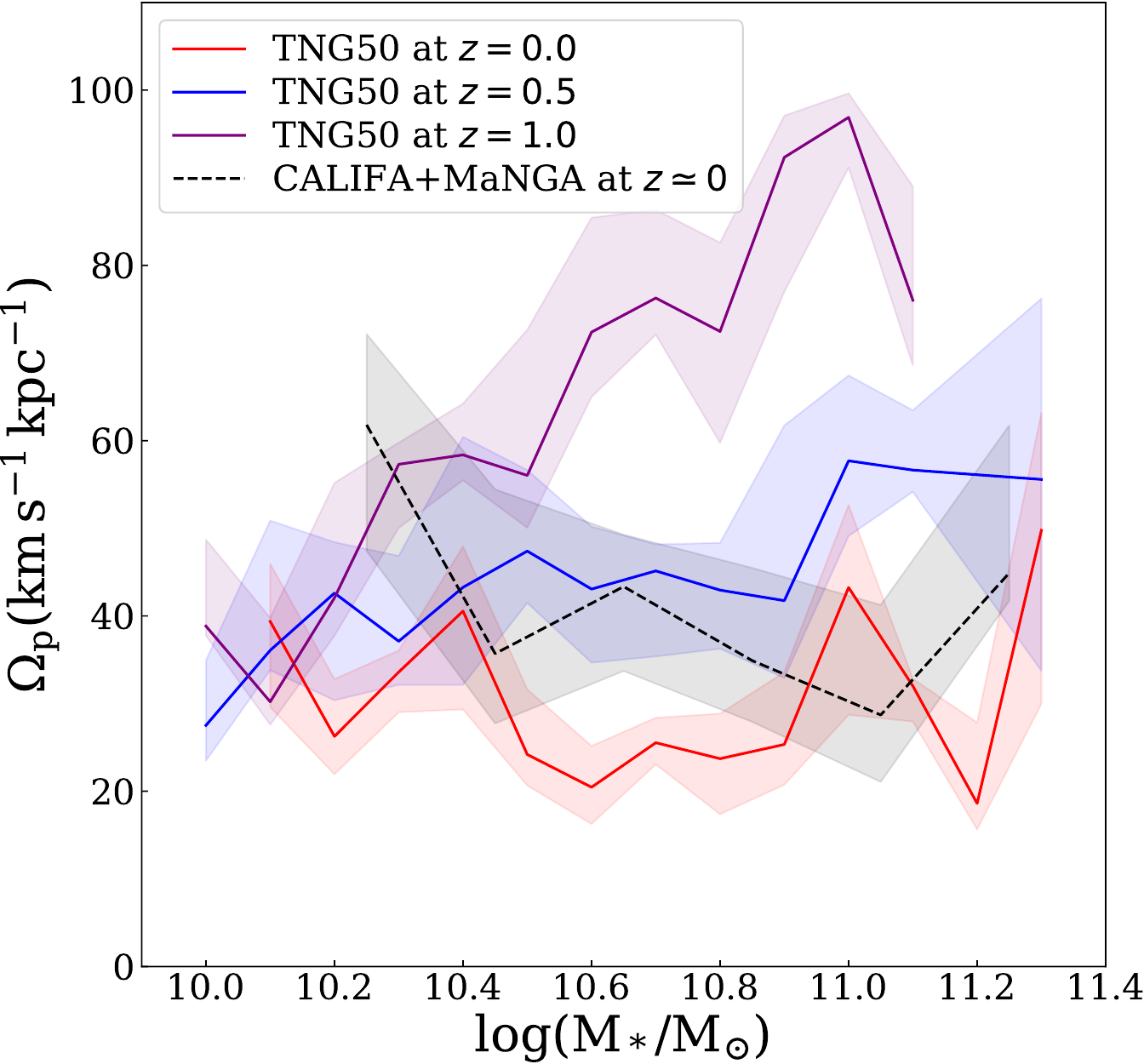}
  \caption{The solid lines display the median value of the bar pattern speed vs stellar mass for TNG50 at different redshifts $z=0.0$, $z=0.5$, and $z=1.0$ and for the observed CALIFA+MaNGA sample at $z\simeq 0.0$. There are 60 galaxies in this sample with $\Delta \Omega_p/\Omega_p \leq 0.5$ that are included in this figure.}
 \label{op_m}
\end{figure}

  \begin{figure}
  \centering
  \includegraphics[width=0.45\textwidth]{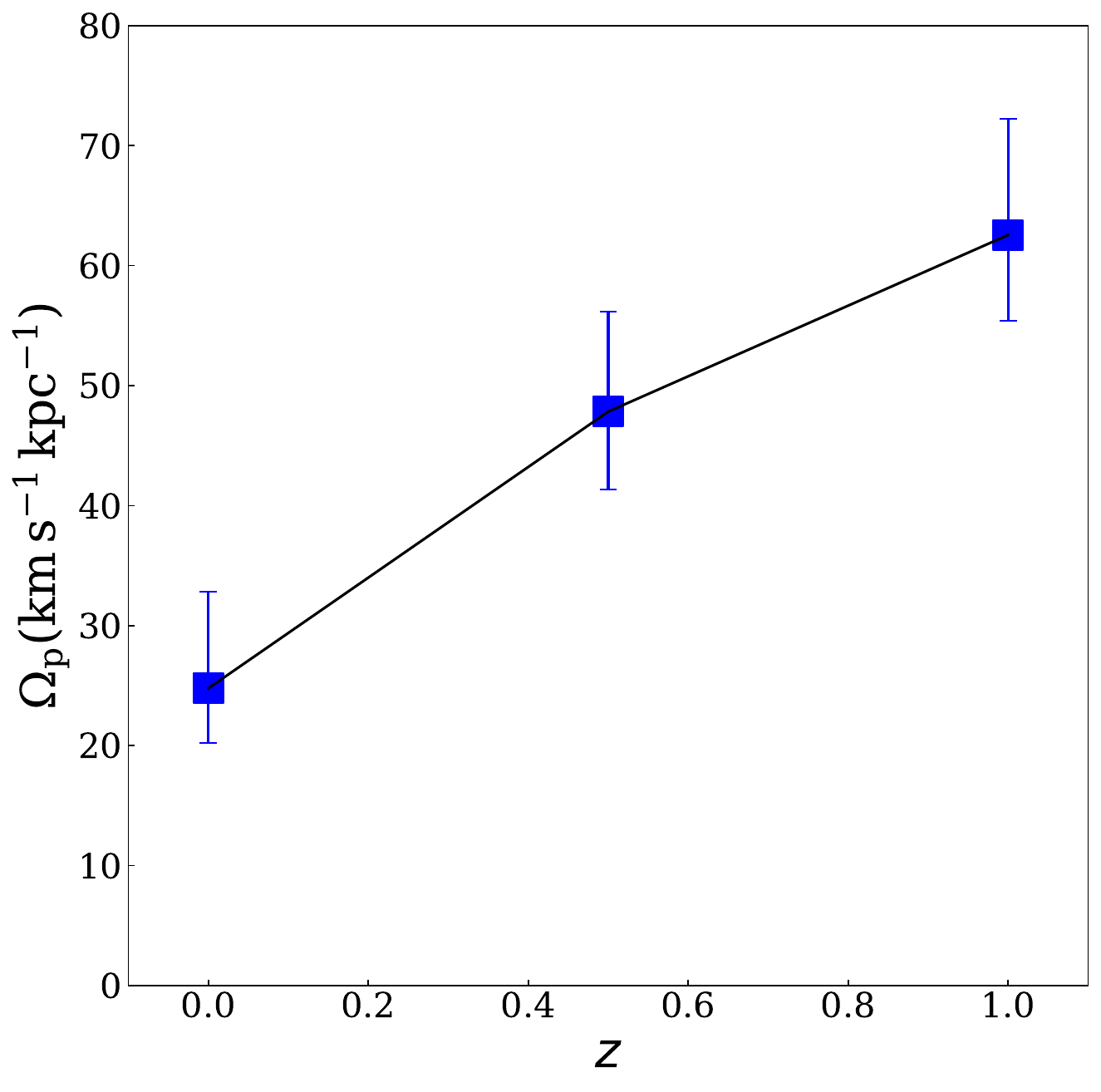}
  \caption{The redshift evolution of the median value of the pattern speed is shown for 59 disk galaxies that remain barred within the redshift interval  $z\leq 0$. The error
bars indicate the $32^{\text{th}}$- $68^{\text{th}}$ percentile.}
 \label{op_new}
\end{figure}

  \begin{figure}
  \centering
  \includegraphics[width=0.45\textwidth]{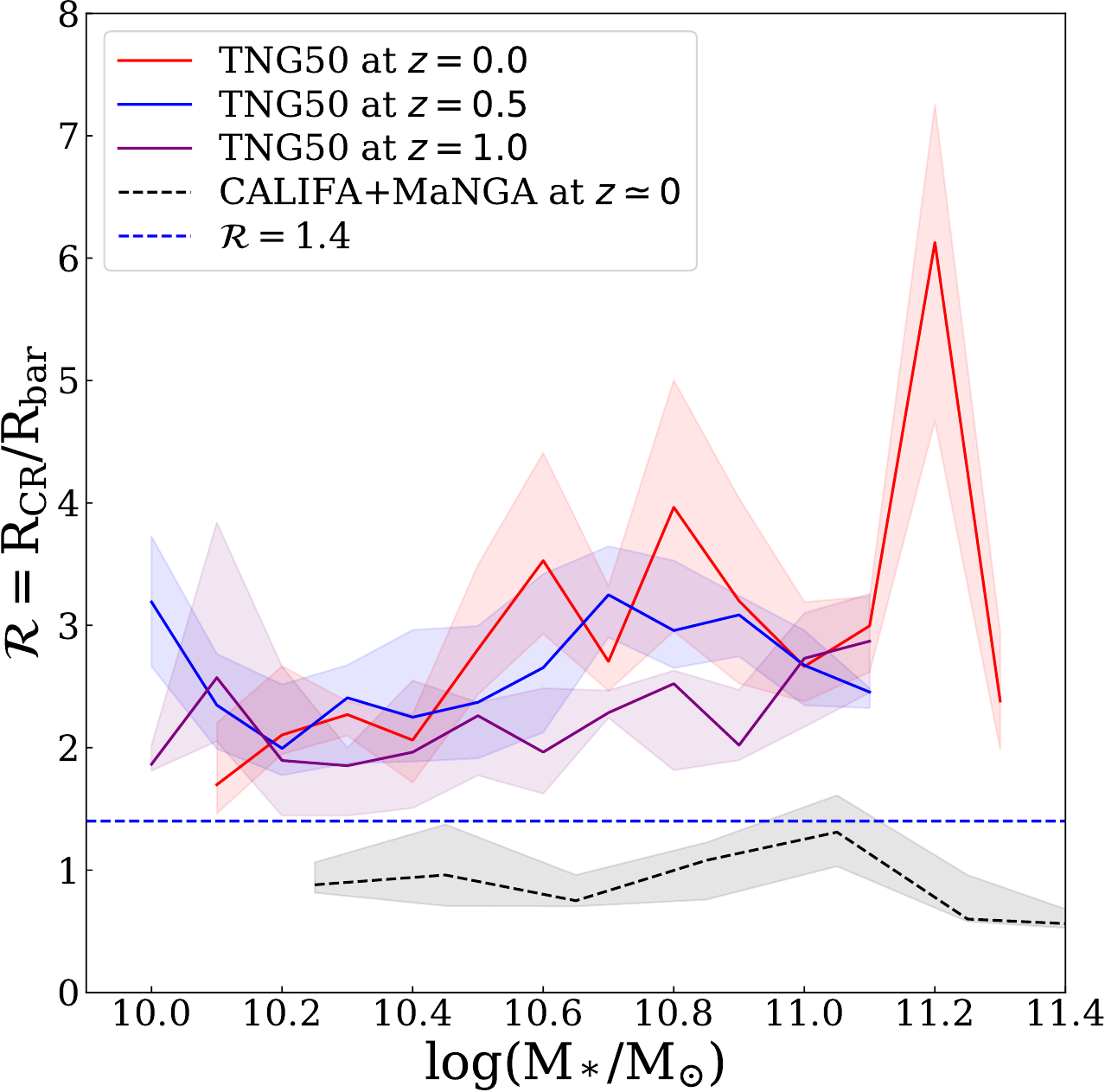}
  \caption{The solid lines display the median value of the $\mathcal{R}$ parameter vs stellar mass for TNG50 at different redshifts $z=0.0$, $z=0.5$, and $z=1.0$ and for the observed 60 galaxies with $\Delta \Omega_p/\Omega_p \leq 0.5$ in CALIFA+MaNGA sample at $z\simeq 0.0$. }
 \label{R_m}
\end{figure}
\label{Rp}

  \begin{figure*}
  \centering
  \includegraphics[width=0.325\textwidth]{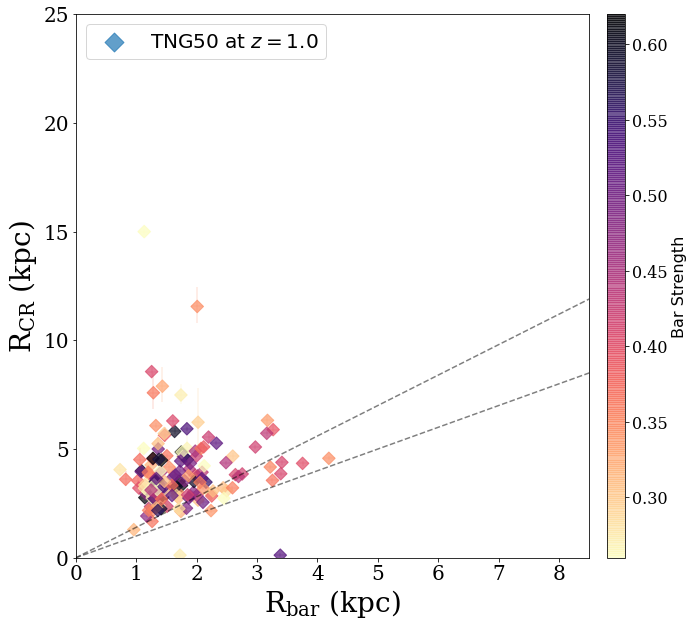}
  \includegraphics[width=0.325\textwidth]{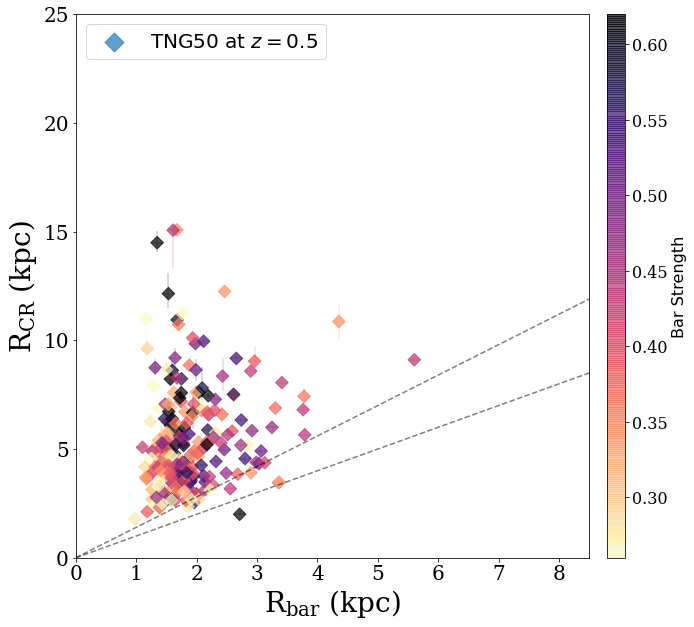}
\includegraphics[width=0.325\textwidth]{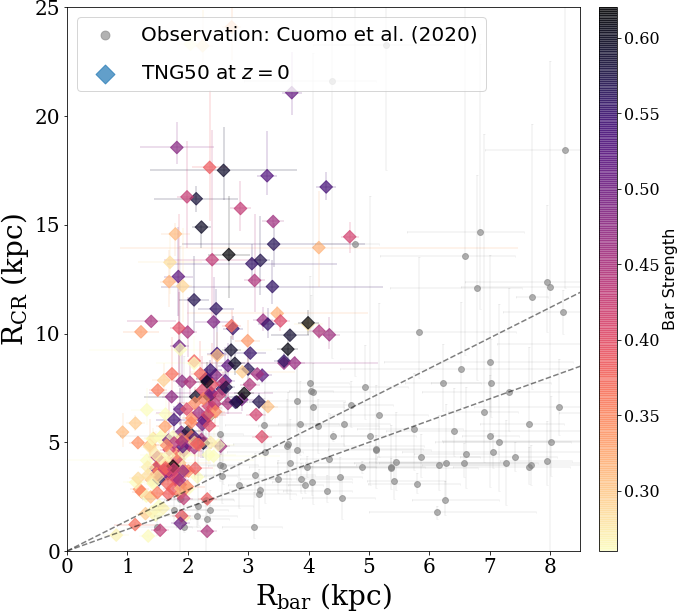}
\caption{Bar corotation radius vs bar length for different redshifts. The color bar corresponds to bar strength $A_2^{max}$. The upper dashed line indicates the fast-slow bar zone. The lower dashed line corresponds to $\mathcal{R}=1$ which indicates the ultrafast bar region. The gray points represent the observation of 60 galaxies with $\Delta \Omega_p/\Omega_p \leq 0.5$ in the CALIFA+MaNGA sample at $z\simeq 0.0$. The right panel bears a resemblance to that of Fig. 5 in the \citet{2021MNRAS.508..926R}. It has been included here to facilitate a more straightforward comparison across the three redshifts.
}
 \label{R_b}
\end{figure*}

The main purpose of this paper is to investigate the redshift evolution of the bar pattern speed. In \cite{2021MNRAS.508..926R} it has been shown that the bars are slow in IllustrisTNG and EAGLE at $z=0.0$ compared to observations. More specifically, the tension between simulation and observations exceeds $5\sigma$. In that work, the origin of this tension was blamed on the existence of dynamical friction caused by dark matter particles. Along the same line, \cite{Algorry2017} and \cite{Peschken2019} report that bars are slow in EAGLE and Illustris, respectively. 

However, only by looking at the single redshift $z=0.0$, it is not possible to conclude that the bars were fast in higher redshifts and got slow due to the dynamical friction. In other words, it might be possible that the pattern speed remains constant with time. In this case, the meaning of the above-mentioned tension would change. To be specific, the parameter $\mathcal{R}=R_{\text{CR}}/R_{\text{bar}}$, the ratio of the corotation radius to the bar radius, is widely used to specify the ``speed'' of the bars. If $\mathcal{R}<1.4$, then the bar is fast, otherwise it is slow. Now, if the pattern speed remains constant with time, and with the restrictive assumption that the rotation curve does not evolve with time, then the corotation radius does not change. In this case, having large values for $\mathcal{R}$ may be directly related to the size of the bars. In other words, the bars can rotate with normal pattern speed $\Omega_p$, while the bar is too short. Therefore the $\mathcal{R}$ parameter would be large, while the dynamical friction has no role. This is the possibility that has been recently raised by  \cite{Frankel2022}. This behaviour is seen for bars in the very high-resolution zoom-in cosmological simulations studied by \cite{Bi2022}. These simulations deal with very gas-rich galaxies. On the other hand, although the $\mathcal{R}$ parameter is large and the bars are classified as being slow, there is no monotonic decline in the bar pattern speed. The bars appear to be too short in these simulations as well. However, In the case of TNG50, as we will show, such behaviour does not appear and the pattern speeds do not stay constant.

It should be stressed that it is essential to understand the origin of the tension. If the bar speed tension is related to the dynamical friction of the dark matter halo, then this would be a serious challenge to the viability of the current dark matter paradigm. This means that either different types of dark matter particles should be postulated \citep{Hui_2016} or new physics is required to address the tension \citep{Roshan2021}. 

Now let us discuss our result. The median value of the pattern speed in terms of stellar mass in different redshifts is illustrated in Fig. \ref{op_m}. On the other hand, the mean value of the pattern speed for all the barred galaxies is given in Table \ref{mean_values}. It is remarkable that the TW method works for 84\%, 88\%, and 77\% of barred galaxies at $z=1.0$, $z=0.5$, and $z=0.0$ respectively, see Table \ref{Sample_sizes}. The TW method does not lead to reliable results for galaxies that host extra features like rings, spirals, or tidal companions. In some cases, we have elongated bulges instead of rotating bars. Therefore, the TW method does not work for them.

We see that at $z=1.0$, the mean pattern speed is around $\bar{\Omega}_p\approx 71\, \text{km}\, \text{s}^{-1}\,\text{kpc}^{-1}$. It decreases by a factor of more than 50\% and reaches $\bar{\Omega}_p\approx 34\, \text{km}\, \text{s}^{-1}\,\text{kpc}^{-1}$ at redshift $z=0.0$. We can see from Fig. \ref{op_m} that at redshift $z=1.0$, the pattern speed is strongly correlated with the stellar mass in the sense that for $M_*\lesssim 10^{11} M_{\odot}$ more massive galaxies have higher pattern speeds. The same behavior is more or less seen in the middle redshift $z=0.5$. However, at $z=0.0$ there is no specific trend. 
This is interesting in the sense that regardless of how fast the bars were in the past, and irrespective of the stellar mass of the host galaxy, they reach almost the same mean value at $z=0.0$. On the other hand, observed galaxies in the local Universe seems to show a weak trend, where low mass galaxies have higher values of $\Omega_p$ (but a slight increase of $\Omega_p$ is observed at high mass regime). This is not observed in the simulated galaxies at $z=0.0$, while the opposite trend is observed in simulated galaxies at $z=1.0$.

{Caution is warranted when interpreting Fig. \ref{op_m} regarding the evolution of pattern speed in relation to stellar mass. The stellar mass of galaxies changes over time, and some galaxies undergo morphological transformations. As a result, the massive disk-shaped galaxies observed at high redshift differ from those observed at low redshift. Therefore, it is essential to focus on a specific sample of galaxies that maintain their disky shape and remain barred from $z=1.0$ to $z=0.0$. This approach allows for a more meaningful quantification of the evolution of pattern speed over time.
In this specific sample, there are 59 galaxies. As the IDs of the galaxies change over time in the TNG simulations, we traced back the merger tree to identify each galaxy's ID at three different redshifts. The median value of the pattern speed for this group is illustrated in Fig. \ref{op_new}, with error bars representing the $32^{\text{th}}$- $68^{\text{th}}$ percentiles. We observe that the median value of the pattern speed increases with redshift, as previously discussed. We will examine the impact of merging on pattern speed in greater detail in Section \ref{merging}.}

It is also interesting to see the behaviour of the $\mathcal{R}$ parameter in terms of stellar mass. The blue dashed line in Fig. \ref{R_m} indicates the border between fast and slow bars, i.e., $\mathcal{R}=1.4$. For $z=1.0$ and $z=0.5$, there is no specific correlation between $\mathcal{R}$ and the stellar mass. However, at $z=0.0$, massive galaxies seem to have higher values of $\mathcal{R}$. The mean value of the $\mathcal{R}$ parameter for all the barred galaxies with reliable pattern speed in each redshift is given in Table \ref{mean_values}. We can see that the mean value increases with time. From this perspective, we confirm the results of \cite{Algorry2017} for the EAGLE simulation. We reiterate that although in TNG50 the bar size increases with time, the corotation radius increases at a higher rate. On the other hand, most of the observed bars at $z=0.0$ are slow. {It is important to note that the large fraction of slow bars at $z=1.0$ does not necessarily imply that these bars have a small pattern speed $\Omega_p$. The length of the bars plays a significant role in determining the magnitude of the $\mathcal{R}$ parameter.}

In Fig. \ref{R_b} we have shown the corotation radius in terms of the bar radius. The color bar indicates the strength of the bars. Each data point corresponds to a barred galaxy. Although at $z=1.0$ all the points are concentrated in the lower left corner of the plot, they expand in the horizontal and vertical direction in the lower redshifts. We can see that the strong bars are longer and also slower in smaller redshifts. This is natural in the sense that strong bars distribute angular momentum more effectively, and consequently experience higher dynamical friction torque. {It is clear that bars increase their length and strength with time, while they decrease their bar pattern speed. }

From an observational standpoint, Fig.~\ref{R_b} confirms that bars observed in the local Universe are slow. The distribution of TNG bars differs from observed bars in the $R_{\text{CR}}$-$R_{\text{bar}}$ plane. While TNG bars cover the same range of values in the $R_{\text{CR}}$ axis as observed bars, they do not replicate the observed bars in the $R_{\text{bar}}$ axis. TNG bars are much smaller in size.

However, the high redshift bar pattern speed observations are poor. So at the moment, it is not possible to compare TNG50's prediction, that bars have much higher $\Omega_p$ at higher redshifts, to the observations. {In particular, no TW measurements of $\Omega_p$ have been performed for galaxies at $z>0.1$ \citep{Cuomo2020}.}

However, \cite{Perez2012} {derived the bar rotation rate from the location of the bar resonances} in a sample of 44 low-inclination ringed galaxies from the SDSS and COSMOS surveys covering the redshift
range $0 < z < 0.8$. Their results imply that there is no evolution for the {bar rotation rate} with redshift. From this perspective, the redshift evolution of the bar pattern speeds in TNG50 seems to conflict with the current observation. However, acquiring more precise observational data is essential to support the discrepancy between observations and simulations regarding the evolution of the bar pattern speed.

Finally, the fast bar tension in $\Lambda$CDM is not just because bars are too short in TNG50, as argued in \cite{Frankel2022}. The problem is twofold in the sense that the corotation radius increases with time as well. We mean that the corotation radius is smaller at higher redshift. Equivalently, the pattern speeds strongly evolve with time in TNG50. This is not consistent with the current observations.

 \subsection{The impact of the gas fraction on pattern speeds}\label{merging}

It would be interesting to discuss the role of the gas fraction in the evolution of the pattern speed. In simulations without gas, the bar length and strength increases and $\mathcal{R}$ goes in the slow regime because of the dynamical friction. In this case, the corotation radius is expected to increase faster than the bar length \citep{Debattista2000}. When gas is present the situation can be much complex. Since gas is dynamically highly responsive, it can readily alter the matter distribution across the disk and influence the evolution of the bar. Several studies have utilized smoothed hydrodynamics simulations to investigate the impact of gas on stellar bars in isolated galaxies \citep{fux99, combes2005, Berentzen2007,Athanassoula2013,seo2019}. It is well-established that the presence of gas weakens the bar, influencing its secular evolution. The gas component may affect the angular momentum transfer by the bar as well. Performing a simulation of a Milky Way-like galactic disk hosting a strong bar, it is shown by \cite{Beane2022} that a small fraction of gas can stabilize the pattern speed and prevent it from slowing down. However, \cite{Beane2022} does not provide any information about the $\mathcal{R}$ parameter. Therefore, it is not possible to infer if the problem is addressed. In other words, the length of the bar and the corotation radius should also be measured and their ratio should be compared with observed galaxies. As already mentioned, \cite{Bi2022} also deals with some gas-rich galactic simulations and find fluctuating pattern speeds that do not slow down with time. However, the $\mathcal{R}$ parameter is high implying that bars are slow

\begin{figure}
  \centering
  \includegraphics[width=0.45\textwidth]{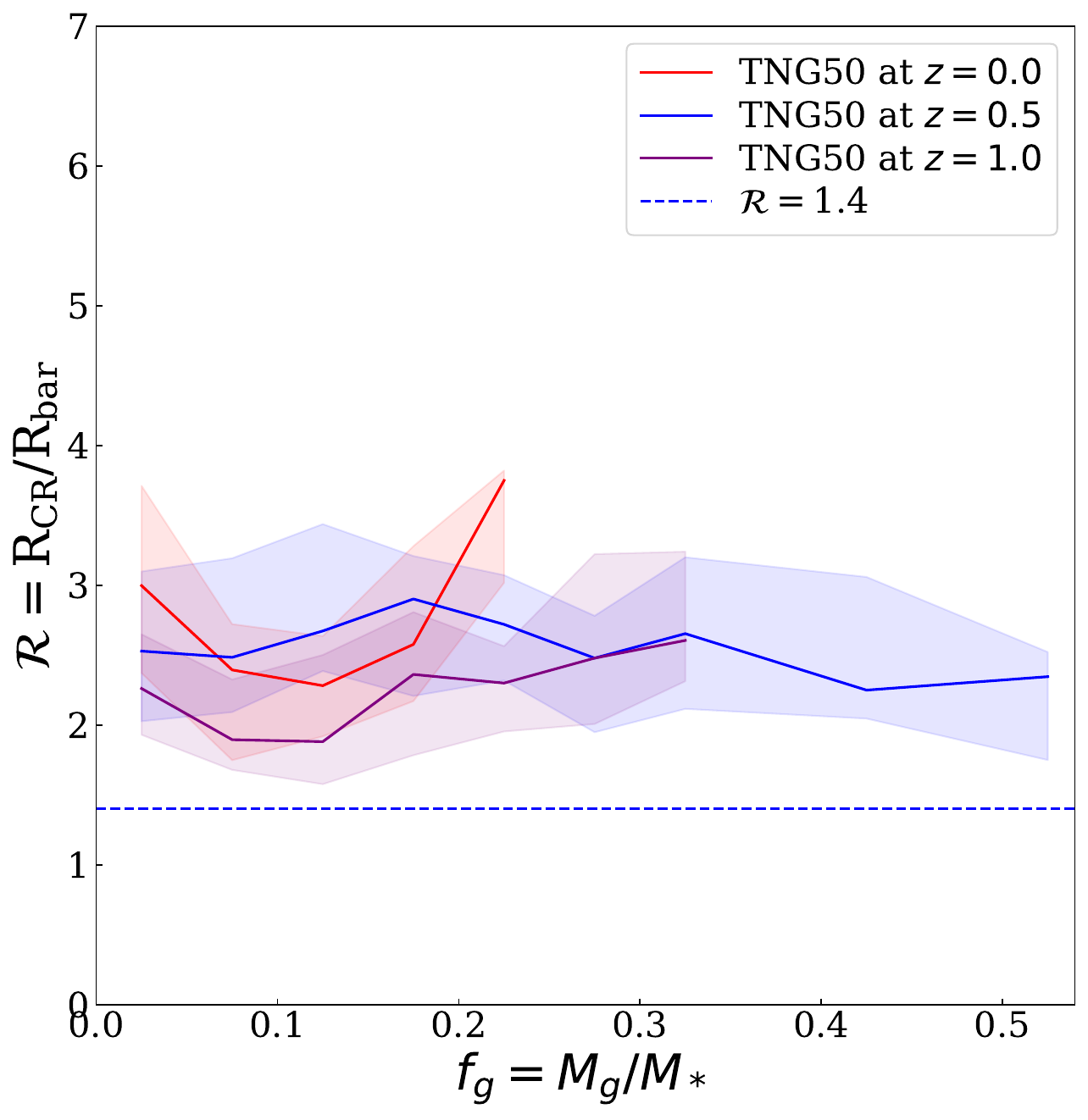}
  \caption{The median value of the $\mathcal{R}$ parameter is illustrated in terms of the gas fraction for $z=0.0$, $z=0.5$ and $z=1.0$. The gas fraction is computed within twice the half-mass radius of the stellar disk.}
 \label{R_g}
\end{figure}

To see the impact of the gas fraction on the bar rotation rate in TNG50, we have plotted the $\mathcal{R}$ parameter in terms of the gas fraction $f_{\text{g}}=M_{\text{g}}/M_*$ in Fig. \ref{R_g} for different redshifts. This fraction is computed within twice the half-mass radius of the stellar disk \citep{2019MNRAS.486.4686K}. If the gas fraction is important, then the $\mathcal{R}$ parameter should be smaller for higher gas fractions. We do not find such a trend in any of the redshifts. In fact, we even observe a rising trend in $\mathcal{R}$ at higher gas fractions in $z=0.0$ and $z=1.0$. Notice that as given in Table \ref{mean_values}, 93\% of the barred galaxies are slow at $z=0.0$. This implies that the presence of gas in the TNG50 simulation does not alleviate the fast bar tension.

As a final remark in this subsection, it is worth mentioning that \cite{Garma2020} found that the $\mathcal{R}$ parameter is positively correlated with the gas fraction. In other words, galaxies with higher gas fraction host bars with higher $\mathcal{R}$ in a {small} sample of galaxies chosen from MaNGA SDSS-IV and CALIFA. A weaker correlation is reported in \cite{2022MNRAS.517.5660G} for {a larger sample of} Milky way analogue galaxies. Anyway, we do not see such a correlation in TNG50. 
 \subsection{The impact of merging on pattern speeds}
   \begin{figure*}
  \centering
    \includegraphics[width=1\textwidth]{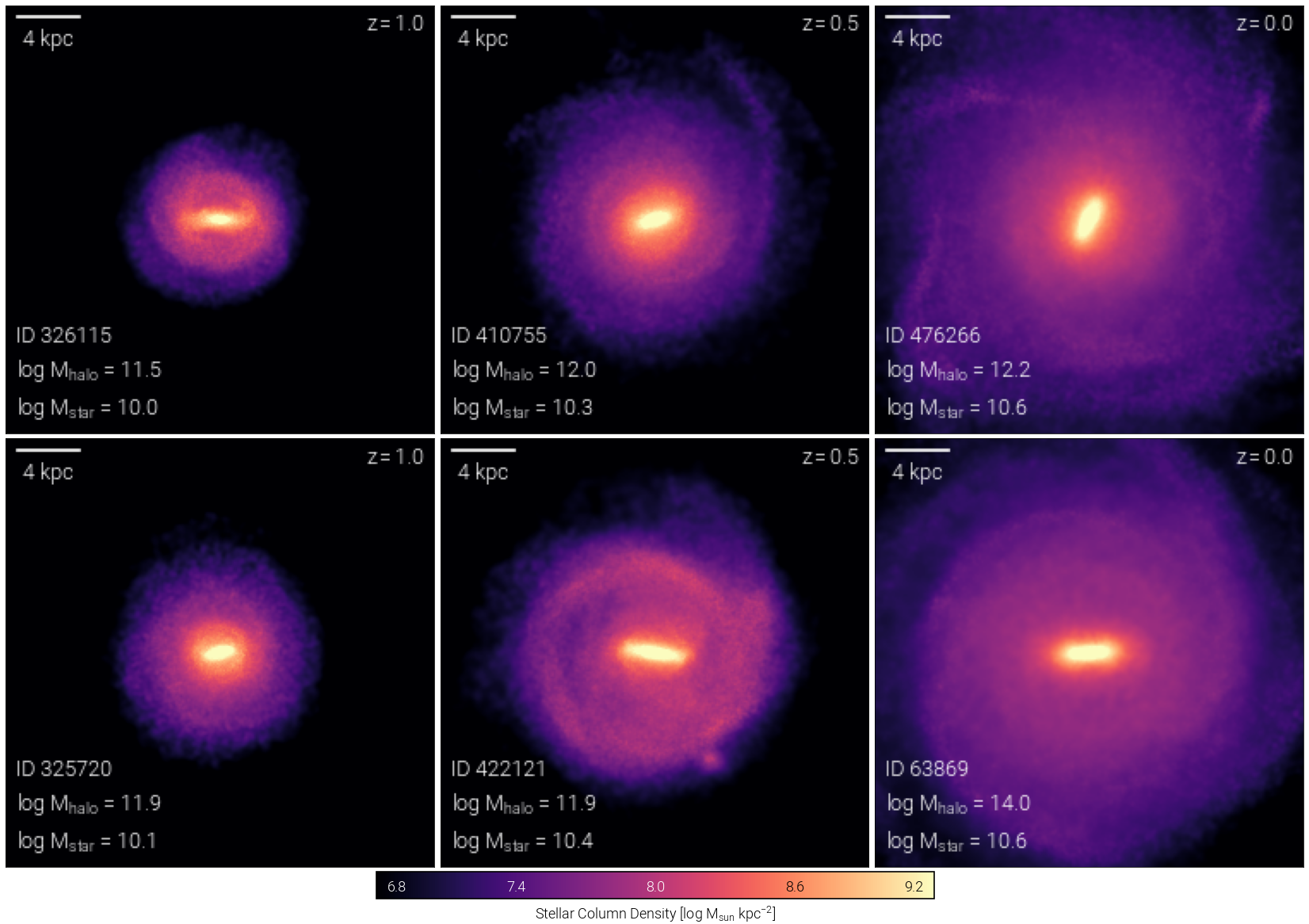}
  %     \includegraphics[width=0.3\textwidth]{z1_major.pdf}
  %   \includegraphics[width=0.3\textwidth]{z5_major.pdf}
  %    \includegraphics[width=0.3\textwidth]{z0_major.pdf}\\
  % \includegraphics[width=0.3\textwidth]{z1.pdf}
  % \includegraphics[width=0.3\textwidth]{z5.pdf}
  % \includegraphics[width=0.3\textwidth]{z0.pdf}
  % % \includegraphics[width=0.3\textwidth]{z1_minor.pdf}
  %  %\includegraphics[width=0.3\textwidth]{z5_minor.pdf}
  %  %\includegraphics[width=0.3\textwidth]{z0_minor.pdf}\\
  \caption{The first row displays the redshift evolution of the stellar component taken from the first group with major mergers. The second row illustrates the evolution of a galaxy from the second group without significant mergers in its history.}
 \label{merging}
\end{figure*}
As already mentioned, although the main reason for decreasing pattern speeds would be the dynamical friction by the dark matter halos, it is necessary to explore the impact of other potential players like merging between the galaxies. Of course, it should be noted that the merging rate between galaxies is influenced by the amount of dynamical friction between them. Anyway, the merger tree of the TNG50 simulation constructed with the \textsc{Sublink} algorithm \citep{Rodriguez2015} can be used to shed light on the significance of the merging. For this purpose, one can walk back in time through the available tree for each galaxy to find when and how many times it has experienced a merger event. As usually done, mergers can be classified according to the stellar mass ratio of the two interacting galaxies at the time of merging. This time is defined as when the secondary galaxy reaches its maximum stellar mass. Here, we divide the barred galaxies at $z=0.0$ that have reliable pattern speed into two groups. The members of the first group have a rich merging history with the stellar mass ratio $> 1/10$ in the redshift interval $z\leq 1$. The members of the second group have no significant merging with the stellar mass ratio $> 1/10$ in redshifts $z\leq 1$. This is done using the catalog provided by \citet{Sotillo2022} and \citet{Eisert2023} where they applied additional checks to clean the trees from non-cosmological cases or spurious flyby and re-merger events. We should also add that we use a version of this catalog in which merging with very small galaxies of no more than 50 stellar particles is ignored. 

There are 17 galaxies in the first group. Interestingly, all these galaxies have a single merger with mass ratio between 1/10 and 1/4 and one more powerful merger with mass ratio $>1/4$. Of course, there are several mergers with mass ratio $<1/10$ for these galaxies. At $z=0.5$, 11 of them are barred, and at $z=1.0$ five galaxies are barred. It is important to note that the presence of bars in these galaxies at $z=0.5$ and $z=1.0$ does not guarantee reliable pattern speeds. Specifically, the number of galaxies with reliable pattern speed is 17, 8 and 4 respectively. The second group which include galaxies with minor merging history, has 183 members. Tracing these galaxies to higher redshifts\footnote{The IDs of the galaxies change over time. For each galaxy at $z=0.0$, we first find its corresponding IDs at higher redshifts and then check its bar properties.} reveals that they are not necessarily barred at higher redshifts. At $z=0.5$, 139 members, and at $z=1.0$ only 86 members are barred. The number of galaxies with reliable pattern speed is 183, 116 and 55 respectively. Identifying galaxies that exhibit bars at all three redshifts and possess reliable pattern speeds would be beneficial. In this case, there are three galaxies within the first group that meet these criteria. The Table \ref{tab2} displays the pattern speeds of these galaxies at various redshifts. Interestingly, the pattern speed of the first two galaxies has notably accelerated from $z=1.0$ to $z=0.0$, contrary to the typical trend where bar pattern speeds tend to decrease over time. By checking the merger tree of these galaxies in more details, we find that the galaxy with ID 96765 has undergone a major merger with the stellar mass ratio $\simeq 0.26$ at $z\simeq 0.18$. Similarly, the second one with ID 476266 has experienced a major merger with the stellar mass ratio $\simeq 0.39$ but at $z\simeq 0.64$. The third galaxy with ID 371126 have a complex merger tree. This galaxy has undergone a major merger with the stellar mass ratio $\simeq 0.38$ at $z=0.33$ and a significant one with the stellar mass ratio $\simeq 1$ at $z=0.15$. A more detailed study is needed to investigate the effect of such mergers. To be specific, let's consider the galaxies within the second group (which have no major mergers) that exhibit bars at all redshifts and have reliable pattern speeds. This subset comprises 52 galaxies. The median pattern speeds for these galaxies are $\bar{\Omega}_p= 62.50^{+6.75}_{-7.15}, 48.84^{+7.41}_{-8.52}, 23.49^{+4.36}_{-5.04} \, \text{km}\, \text{s}^{-1}\,\text{kpc}^{-1}$ for $z=1.0, 0.5$, and $0.0$, respectively. A clear decreasing trend in pattern speeds over time is evident in these galaxies. There is only one galaxy (ID: 229935) in this subset that its pattern speed has significantly increased form $\bar{\Omega}_p\simeq 25.9\, \text{km}\, \text{s}^{-1}\,\text{kpc}^{-1}$ at $z=1.0$ to $\bar{\Omega}_p\simeq 41.2 \, \text{km}\, \text{s}^{-1}\,\text{kpc}^{-1}$ at $z=0.0$.

The analysis above highlights the impact of galaxy mergers on bar pattern speeds. It is observed that two out of three galaxies involved in major mergers deviate from the typical pattern speed evolution trend. However, it still can be inferred that mergers are not the primary factor contributing to the presence of slow bars in TNG50. This conclusion is supported by two key points: firstly, mergers can lead to an increase in the pattern speeds of galaxies; secondly, the majority of galaxies in our overall sample do not undergo major mergers within the redshift range $z\leq 1$. Specifically, within the second group, galaxies tend to exhibit slower patterns at $z=0.0$, despite lacking major mergers in their merger history. 

\begin{table}
	\centering
	\begin{tabular}{ccccc}
		\hline
		\multirow{2}{*}{} Galaxy ID& $\Omega_p(z=0)$ & $\Omega_p(z=0.5)$ & $\Omega_p(z=1.0)$ \\
		\hline
		96765 & 77.22$\pm$ 0.77 & 32.59 $\pm$ 0.95 & 32.04 $\pm$ 1.37  \\
		476266& 89.30 $\pm$ 0.50 & 43.26$\pm$1.74 & 28.03$\pm$1.24  \\
		371126 & 36.54 $\pm$ 0.84 & 44.08$\pm$0.61 & 76.55$\pm$0.51  \\
	
		\hline
	\end{tabular}
	\caption{The pattern speeds for the three galaxies 96765, 476266 and 371126 which exhibit bar at all three redshifts and undergo major mergers within the redshift range $z\leq 1$. The pattern speed is measured in $\text{km}\, \text{s}^{-1}\,\text{kpc}^{-1}$. }
	\label{tab2}
\end{table}

It is useful to see the time evolution of a single galaxy from each group. The top row in Fig. \ref{merging} shows the redshift evolution of the stellar component of a galaxy with ID 476266 (at $z=0.0$) representing the first group characterized by major mergers. The radial expansion of the disk aligns with the findings of \cite{Villalba2022}, demonstrating that the disk galaxies' scale length increases over time in TNG50 galaxies. As detailed in Table \ref{tab2}, the pattern speed of this galaxy rises by approximately 319\% from $z=1.0$ to $z=0.0$. This atypical behavior may be directly linked to the significant mergers experienced by the galaxy.

On the other hand, the second row in Fig. \ref{merging} shows the redshift evolution of the stellar component of a galaxy with ID 63869 (at $z=0.0$). This galaxy has been illustrated as a representative of the second group. In addition to the radial expansion, it is evident that the length of the bar increases over time. The pattern speed of this galaxy is $\Omega_p\simeq 47.48$, 43.54, and 25.02 $\text{km}\,\text{s}^{-1}\,\text{kpc}^{-1}$ at $z=1.0$, 0.5, and 0.0 respectively. Therefore the pattern speed decreases by a factor of 47\% from $z=1.0$ to $z=0.0$, whereas the galaxy does not undergo any significant merging. So the reduction in the pattern speed of the bar is due to the secular evolution of the disk. One may think that the bar speed decreases due to the angular momentum transfer by the bar to the outer parts of the disk. This may seem sensible since the disk expands in the radial direction. However, it is interesting to mention that according to \cite{Villalba2022}, the radial scale length of the disks grows with time in TNG50 at almost the same rate for barred and unbarred disks (even slower for barred galaxies), see Fig. 3 in \cite{Villalba2022}. Furthermore, the paper suggests that unbarred galaxies in TNG50 are radially more extended compared to barred ones. This directly means that the radial expansion is not merely due to the angular momentum loss by the stellar bar. On the other hand, we show that the angular momentum exchange between the bar and dark matter halo is effectively happening and is the main reason for decreasing pattern speed.

\begin{figure}
  \centering
  \includegraphics[width=0.45\textwidth]{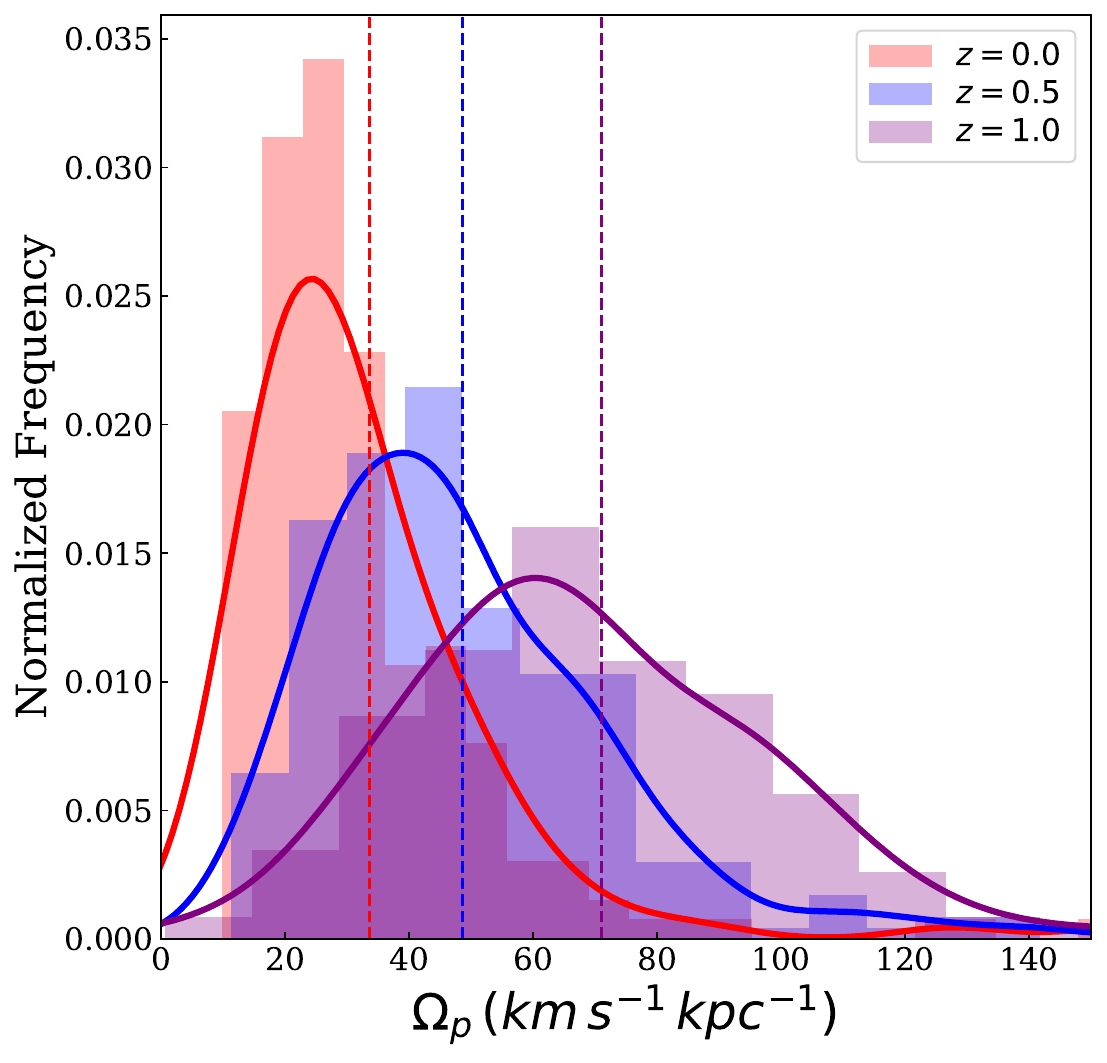}
  \caption{The normalized histogram of the pattern speed at different $z$: the vertical dashed lines show the median values of the pattern speed for every histogram, while the full lines are kernel density estimates of these histograms at different redshifts.}
 \label{hist}
\end{figure}

\section{discussion and conclusion }
\label{conclusion}
\begin{figure}
  \centering
  \includegraphics[width=0.45\textwidth]{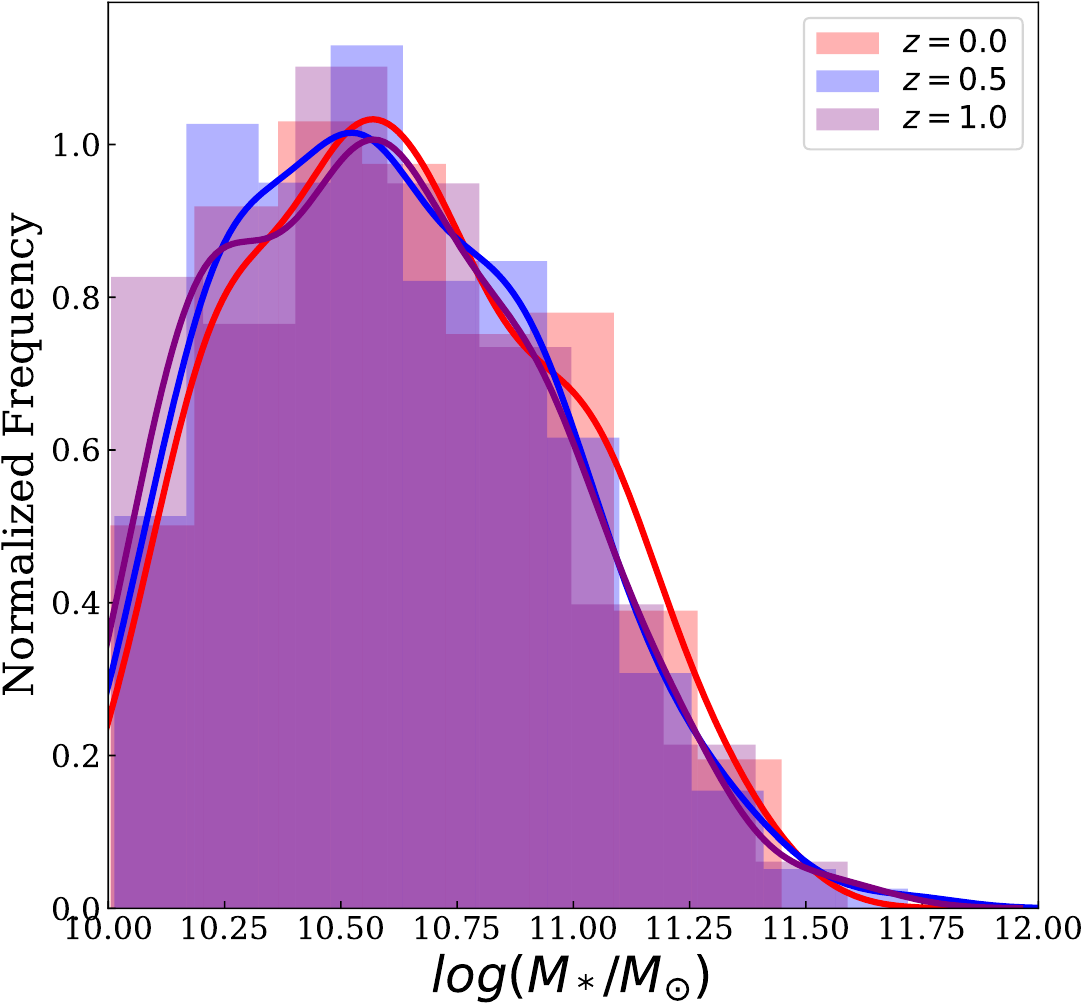}
  \caption{The distribution of the stellar mass in different redshifts. It is seen there is no bias and the same range of masses are considered in different redshifts. }
 \label{mass_dis}
\end{figure}
This paper investigates the redshift evolution of the bar fraction, bar length, and pattern speed of TNG50 galaxies. Regarding the bar fraction, we confirm the already known problem that cosmological simulations lack bars in less massive galaxies. Furthermore, we showed that the bar fraction increases with redshift for galaxies in the stellar mass range $M_*\gtrsim 10^{10}\, M_{\odot}$. This also seems inconsistent with the observations that imply the opposite.  

Regarding the bar length, we found two specific features: i) the median value of the bar length decreases with redshift. ii) there is no specific correlation between the bar length and stellar mass in a given redshift. The comparison of both features with relevant observations reveals a conflict with the observations. To assess the second feature, we utilized a sample of galaxies from the MaNGA and CALIFA surveys, and demonstrated a clear discrepancy between bar lengths in TNG50 and observations. We also discussed similar results that have been previously reported in the literature.

The pattern speed evolution in TNG50 is the main purpose of this paper. The median value of the pattern speed changes with redshift. To be specific, the mean pattern speed is around $\bar{\Omega}_p\approx 71\, \text{km}\, \text{s}^{-1}\,\text{kpc}^{-1}$ at $z=1.0$. It decreases by a factor of more than 50\% and reaches $\bar{\Omega}_p\approx 34\, \text{km}\, \text{s}^{-1}\,\text{kpc}^{-1}$ at redshift $z=0.0$. The distribution of $\Omega_p$ at different redshifts is illustrated in Fig. \ref{hist}. It is seen that the median value of $\Omega_p$ decreases with time. It is worth noting that there are only two galaxies with $\Omega_p < 9.7\,\text{km}\, \text{s}^{-1}\,\text{kpc}^{-1}$ in the entire sample of galaxies with masses $M_*\geq 10^{10} M_{\odot}$ at all three redshifts.

In this study, we analyze three galaxy populations at redshifts $z=0.0, 0.5$, and $1.0$ to compare their bar pattern characteristics, mirroring observational practices. Our focus is not on tracking the evolution of individual galaxies. Fig. \ref{mass_dis} illustrates that these populations exhibit similar mass distributions, indicating no bias in mass across the three groups.

To investigate the impact of galaxy mergers on the deceleration of pattern speed, we categorized TNG50 galaxies into two groups. The first group comprises galaxies which have undergone major mergers in their history, while the second group consists of galaxies with no significant mergers in their merger tree. Among these groups, only three galaxies in the first group exhibit bar structures at all three redshifts and possess reliable pattern speeds. Notably, for two of these galaxies, the pattern speed experiences a notable increase over time, indicating that major mergers can amplify the rotational speed of galactic bars. Conversely, the second group includes 52 galaxies with barred structures at $z\leq 1$ and reliable pattern speeds, where the average pattern speed decreases over time. This suggests that the majority of the galaxies in our entire sample do not undergo major mergers within the redshift range $z\leq 1$, implying that the decline in average pattern speed is not linked to intergalactic mergers. Furthermore, we studied the role of the gas fraction. We showed that the $\mathcal{R}$ parameter does not correlate with the gas fraction. This means that in TNG50, the existence of gas does not alleviate the fast bar tension.

Our results reveal a clear prediction by the standard cosmological model that the mean pattern speed of barred galaxies must be much higher in high redshifts $(z\sim 1)$. Currently, since the pattern speed observations are limited to nearby galaxies, it is not possible to check the viability of this prediction. Future high redshift observations would shed light on this issue.

\section*{Acknowledgements}
We appreciate the anonymous referee for constructive comments. We appreciate Tahere Kashfi for providing us with the gas fraction data in TNG50 galaxies at $z=0.0$. The work of Asiyeh Habibi, Mahmood Roshan, and Sharam Abbassi is supported by the Ferdowsi University of Mashhad. The main part of the calculation
was conducted at the Sci-HPC center of the Ferdowsi University of Mashhad. The research of J. Alfonso L. Aguerri has been supported by the Spanish Ministry of Ciencia e Innovación under the grant PID2020-119342GB-I00. Virginia Cuomo acknowledges the support provided by Chilean ANID through 2022 FONDECYT postdoctoral research grant no. 3220206.

%\section*{DATA AVAILABILITY}
%The simulation data used in this paper are publicly available on the IllustrisTNG website. The pattern speed observations data implemented in Fig. \ref{R_b} are available in \cite{Cuomo2020}. All the data generated in this paper (e.g. pattern speeds, corotation radii, bar lengths, etc at different redshifts) are available upon reasonable request to AH.

\bibliographystyle{aa}
\bibliography{short,HZ_PS}
\end{document}